\documentclass[11pt,twoside,letterpaper]{article} 
\usepackage{amssymb}
\begin{document}
\title{Quark confinement mechanism for baryons}
\author{\bfseries\itshape Yu. P. Goncharov
\thanks{E-mail address: 
ygonch@chern.hop.stu.neva.ru}\\
Theoretical Group, Experimental Physics Department, State\\
Polytechnical University,
Sankt-Petersburg 195251, Russia}
\date{}
\maketitle
{\begin{flushleft}
\vskip 0.45in
\abstract{
The confinement mechanism proposed earlier and then successfully applied 
to meson spectroscopy by the author is extended over baryons.
For this aim the wave functions of baryons are built as tensorial products 
of those corresponding to the 2-body problem underlying the confinement 
mechanism of two quarks. This allows one to obtain the Hamiltonian of the 
quark interactions in a baryon and, accordingly, the possible energy spectrum 
of the latter. Also one may 
construct the electric and magnetic form factors of baryon in a natural way 
which entails the expressions for the root-mean-square radius and anomalous
magnetic moment. To ullustrate the formalism in the given Chapter for the sake of 
simplicity only symmetrical baryons (i.e., composed from three quarks of the 
same flavours) $\Delta^{++}$, 
$\Delta^{-}$, $\Omega^-$ are considered. For them 
the masses, the root-mean-square radii and anomalous magnetic moments are 
expressed in an explicit analytical form through the parameters of the confining 
SU(3)-gluonic field among quarks and that enables one to get a number of numerical 
estimates for the mentioned parameters from experimental data. We also discuss chiral 
limit for the baryons under consideration and estimate the purely gluonic 
contribution to their masses. Further the problem of masses in particle physics is
shortly discussed within the framework of the given approach. Finally, a few remarks 
are made about the so-called Yang-Mills Millennium problem and a possible way for 
proving it is outlined. 
} 
\end{flushleft}
\vspace{1.0cm}

\vspace{.08in} 


\tableofcontents
\section{Introduction}
The present Chapter is a natural continuation of our previous papers on studying 
both heavy quarkonia and the pseudoscalar meson nonet within the framework of 
the (quark) confinement 
mechanism proposed earlier by the author. Global strategy may consist in reconsidering the whole 
spectroscopy of both mesons and baryons from the positions of the mentioned 
mechanism so, e.g., exploring the pseudoscalar meson nonet is in essence only the 
first step in the given direction. But an extension of the approach over 
baryons requires a relativistic 3-body problem to be considered 
in contrast to mesons where we deal with a relativistic 2-body problem 
as follows from the standard quark model (SQM). It is clear, however, that 
the confinement mechanism under discussion (which in essence describes the 
mentioned relativistic 2-body problem) should also occupy a fitting place in 
the 3-body constructions. Let us, therefore, shortly outline the main features 
of the approach suggested. 

In \cite{{Gon01},{Gon051},{Gon052}} for the Dirac-Yang-Mills 
system derived from 
QCD-Lagrangian  an unique family of compatible 
nonperturbative solutions was found and explored, which could pretend to 
decsribing confinement of two quarks. 
The applications of the family to the description of both the heavy quarkonia 
spectra \cite{{Gon031},{Gon032},{Gon04},{Gon08a}} and a number of properties of 
pions, kaons, $\eta$- and $\eta^\prime$-mesons 
\cite{{Gon06},{Gon07a},{Gon07b},{Gon08},{Gon08b},{Gon10}} showed that the 
confinement mechanism is qualitatively the same for both light mesons and 
heavy quarkonia.
At this moment it can be decribed in the following way.

The next main physical reasons underlie linear confinement in the 
mechanism under discussion. The first one is that gluon exchange between 
quarks is realized with the propagator different from the photon-like one, and 
existence and form of such a propagator is a {\em direct} consequence of the 
unique confining 
nonperturbative solutions of the Yang-Mills equations 
\cite{{Gon051},{Gon052}}. The second reason is that, 
owing to the structure of the mentioned propagator, quarks mainly emit and 
interchange the soft gluons so the gluon condensate (a classical gluon field) 
between quarks basically consists of soft gluons (for more details 
see Refs. \cite{{Gon051},{Gon052}}) but, because of the fact that any gluon 
also emits gluons (still softer), the corresponding gluon concentrations 
rapidly become huge and form a linear confining magnetic colour field of 
enormous strengths, which leads to confinement of quarks. This is by virtue of 
the fact that just the magnetic part of the mentioned propagator is responsible 
for a larger portion of gluon concentrations at large distances since the 
magnetic part has stronger infrared singularities than the electric one. 
In the circumstances 
physically nonlinearity of the Yang-Mills equations effectively vanishes so the 
latter possess the unique nonperturbative confining solutions of the 
Abelian-like form (with the values in Cartan subalgebra of SU(3)-Lie algebra) 
\cite{{Gon051},{Gon052}} which describe the gluon condensate under 
consideration. Moreover, since the overwhelming 
majority of gluons is soft they cannot leave the hadron (meson) until some 
gluons obtain additional energy (due to an external reason) to rush out. So 
we also deal with the confinement of gluons.  

The approach under discussion equips us with the explicit wave functions 
for every two quarks (meson or quarkonium). The wave functions are parametrized 
by a set of 
real constants $a_j, b_j, B_j$ describing the mentioned 
{\em nonperturbative} confining SU(3)-gluonic field (the gluon condensate) and 
they are {\em nonperturbative} modulo square integrable 
solutions of the Dirac equation in the above confining SU(3)-field and also  
depend on $\mu_0$, the reduced
mass of the current masses of quarks forming meson. It is clear that under the 
given approach just constants $a_j, b_j, B_j,\mu_0$ determine all properties 
of any meson (quarkonium), i. e.,  the approach directly appeals to quark 
and gluonic degrees of freedom as should be according to the first principles 
of QCD. Also it is clear that the constants mentioned should be extracted from 
experimental data. 

Such a program has been to a certain extent advanced in 
Refs. 
\cite{{Gon031},{Gon032},{Gon04},{Gon08a},{Gon06},{Gon07a},{Gon07b},{Gon08},{Gon08b},{Gon10}}  
for both heavy 
quarkonia and the pseudoscalar meson nonet. It is clear, however, that baryons 
constitute no less physically important class of hadrons. But according to 
SQM (see, e. g., \cite{pdg}) baryons are composed from three quarks, generally 
speaking, with different flavours. As a consequence, we are faced with a 
relativistic 3-body problem. 

Under the circumstances the main aim of the present Chapter is to develop a 
certain approach to baryons based on the above confinement mechanism. For the 
sake of simplicity we shall here restrict ourselves to the symmetrical baryons 
(i.e., composed from three quarks of the same flavours) $\Delta^{++}$, 
$\Delta^{-}$, $\Omega^-$ inasmuch as in accordance with SQM $\Delta^{++}=uuu$, 
$\Delta^{-}=ddd$, $\Omega^-=sss$.

Section 2 contains a survey of main relations underlying 
description of any mesons (quarkonia) in our approach (the 2-body problem). 
Section 3 takes into account the considerations of Section 2 when constructing 
possible wave functions of baryons in our approach. Section 4 uses the obtained 
baryonic wave functions to build masses, the electric form factors, the 
root-mean-square radii $<R>$ and the anomalous magnetic moments of the 
symmetrical baryons under consideration in an explicit analytic form. Section 
5 gives estimates for parameters of the confining SU(3)-gluonic field for 
baryons under discussion while Section 6 deals with discussion about chiral 
limit for baryons under studying. In Section 7 the problem of masses in particle 
physics is shortly discussed within the framework of approach to the chiral symmetry 
breaking in QCD proposed recently by the author. In Section 8 we make a few remarks 
about the so-called Yang-Mills Millennium problem and outline a possible way for 
proving it. Section 9 is devoted to a number of concluding remarks. 

 Appendices A and B contain the detailed description 
of main building blocks for meson wave functions in the approach under 
discussion, respectively: eigenspinors of the Euclidean Dirac operator on 
2-sphere ${\mathbb S}^2$ and radial parts for the modulo square integrable 
solutions of Dirac equation in the confining SU(3)-Yang-Mills field while 
Appendix C recalls a few facts about tensorial products of Hilbert spaces 
necessary to construct the baryonic wave functions. At last, 
Appendix D supplements Section 2 with a proof of the uniqueness theorem from 
that Section in the case of SU(3)-Yang-Mills equations.
 
Further we shall deal with the metric of
the flat Minkowski spacetime $M$ that
we write down (using the ordinary set of local spherical coordinates
$r,\vartheta,\varphi$ for the spatial part) in the form
$$ds^2=g_{\mu\nu}dx^\mu\otimes dx^\nu\equiv
dt^2-dr^2-r^2(d\vartheta^2+\sin^2\vartheta d\varphi^2)\>. \eqno(1)$$
Besides, we have $|\delta|=|\det(g_{\mu\nu})|=(r^2\sin\vartheta)^2$
and $0\leq r<\infty$, $0\leq\vartheta<\pi$,
$0\leq\varphi<2\pi$.

Throughout the Chapter we employ the Heaviside-Lorentz system of units 
with $\hbar=c=1$, unless explicitly stated otherwise, so the gauge coupling 
constant $g$ and the strong coupling constant ${\alpha_s}$ are connected by 
the relation $g^2/(4\pi)=\alpha_s$. 
Further, we shall denote by $L_2(F)$ the set of the modulo square integrable
complex functions on any manifold $F$ furnished with an integration measure, 
then $L^n_2(F)$ will be the $n$-fold direct product of $L_2(F)$
endowed with the obvious scalar product while $\dag$ and $\ast$ stand, 
respectively, for Hermitian and complex conjugation. Our choice of Dirac 
$\gamma$-matrices conforms to the so-called standard representation and is 
the same as in Ref. \cite{Gon06}. At last $\otimes$ means 
tensorial product of matrices and $I_n$ is the unit $n\times n$ matrix so that, 
e.g., we have 
$$I_3\otimes\gamma^\mu=
\pmatrix{\gamma^\mu&0&0\cr 0&\gamma^\mu&0\cr 0&0&\gamma^\mu\cr}$$ 
for any Dirac $\gamma$-matrix $\gamma^\mu$ and so forth. 

When calculating we apply the 
relations $1\ {\rm GeV^{-1}}\approx0.1973269679\ {\rm fm}\>$,
$1\ {\rm s^{-1}}\approx0.658211915\times10^{-24}\ {\rm GeV}\>$, 
$1\ {\rm V/m}\approx0.2309956375\times 10^{-23}\ {\rm GeV}^2$, 
$1\ {\rm T}=4\pi\times10^{-7} {\rm H/m}\times1\ {\rm A/m}
\approx0.6925075988\times 10^{-15}\ {\rm GeV}^2 $, $\mu_N=
3.1524512326\times10^{-17}\ {\rm GeV}/{\rm T}
\approx0.4552226197\times10^{-4}\ {\rm GeV}^{-1}$.  

Finally, for the necessary estimates we shall employ the $T_{00}$-component 
(volumetric energy density ) of the energy-momentum tensor for a 
SU(3)-Yang-Mills field which should be written in the chosen system of units 
in the form
$$T_{\mu\nu}=-F^a_{\mu\alpha}\,F^a_{\nu\beta}\,g^{\alpha\beta}+
{1\over4}F^a_{\beta\gamma}\,F^a_{\alpha\delta}g^{\alpha\beta}g^{\gamma\delta}
g_{\mu\nu}\>. \eqno(2) $$

\section{Survey of main relations for the 2-body problem}
\subsection{Preliminaries}
As was mentioned above, our considerations shall be based on the unique family 
of compatible nonperturbative solutions for 
the Dirac-Yang-Mills system (derived from QCD-Lagrangian) studied at the whole 
length in Refs. \cite{{Gon01},{Gon051},{Gon052}}.  Referring for more details 
to those references, let us briefly decribe and specify only the relations 
necessary to us in the present necessary to us in the present necessary to us 
in the present Chapter. 

One part of the mentioned family is presented by the unique nonperturbative 
confining solution of the SU(3)-Yang-Mills 
equations for the gluonic field $A=A_\mu dx^\mu=
A^a_\mu \lambda_adx^\mu$ ($\lambda_a$ are the 
known Gell-Mann matrices, $\mu=t,r,\vartheta,\varphi$, $a=1,...,8$) and looks 
as follows 
$$ {\mathcal A}_{1t}\equiv A^3_t+\frac{1}{\sqrt{3}}A^8_t =
-\frac{a_1}{r}+A_1 \>,$$
$${\mathcal A}_{2t}\equiv -A^3_t+\frac{1}{\sqrt{3}}A^8_t=
-\frac{a_2}{r}+A_2\>,$$
$${\mathcal A}_{3t}\equiv-\frac{2}{\sqrt{3}}A^8_t=
\frac{a_1+a_2}{r}-(A_1+A_2)\>, $$
$${\mathcal A}_{1\varphi}\equiv A^3_\varphi+\frac{1}{\sqrt{3}}A^8_\varphi=
b_1r+B_1 \>,$$
$${\mathcal A}_{2\varphi}\equiv -A^3_\varphi+\frac{1}{\sqrt{3}}A^8_\varphi=
b_2r+B_2\>,$$
$${\mathcal A}_{3\varphi}\equiv-\frac{2}{\sqrt{3}}A^8_\varphi=
-(b_1+b_2)r-(B_1+B_2)\> \eqno(3)$$

with the real constants $a_j, A_j, b_j, B_j$ parametrizing the family. 
The word {\em unique} should be understood in the strict mathematical sense. 
In fact in Ref. \cite{Gon051} the following theorem was proved 
(see also appendix D):

{\em The unique exact spherically symmetric (nonperturbative) solutions 
(depending only on $r$ and $r^{-1}$) of SU(3)-Yang-Mills equations in Minkowski spacetime 
consist of the family of (3)}. 

Additionally we impose the Lorentz condition on the sought solutions. 
The latter condition is necessary for 
quantizing the gauge fields consistently within the framework of perturbation 
theory (see, e. g. Ref. \cite{Ryd85}), so we should impose the given condition 
that can be written
in the form ${\rm div}(A)=0$, where the divergence of the Lie algebra valued
1-form $A=A_\mu dx^\mu=A^a_\mu \lambda_adx^\mu$ is defined by the relation 
(see, e. g., Refs. \cite{81})
$${\mathrm{div}}\, {A}=\ast(d\ast{A})=
\frac{1}{\sqrt{\delta}}\partial_\mu(\sqrt{\delta}g^{\mu\nu}
A_\nu)\>$$
with the Hodge star operator $\ast$ (see Appendix D).
It should be emphasized that, from the physical point of view, the Lorentz 
condition reflects the fact of transversality for gluons that arise as quanta 
of SU(3)-Yang-Mills field when quantizing the latter (see, e. g., 
Ref. \cite{Ryd85}). 

It should be noted that solution (3) was found early in 
Ref. \cite{Gon01} but its uniqueness was proved just in Ref. \cite{Gon051} 
(see also Ref. \cite{Gon052}). Besides, in Ref. \cite{Gon051} (see also 
Ref. \cite{Gon06}) it was shown that the above unique confining solutions (3) 
satisfy the so-called Wilson confinement criterion \cite{Wil}. Up to now 
nobody contested the above results so if we want to describe interaction 
between quarks by spherically symmetric SU(3)-fields then they can be only 
those from the above theorem. On the other hand, the desirability of 
spherically symmetric (colour) interaction between quarks at all distances 
naturally follows from analysing the $p\bar{p}$-collisions (see, e.g., 
Ref. \cite{Per}) where one observes a Coulomb-like potential in events which 
can be identified with scattering quarks on each other, i.e., actually at small 
distances one observes the Coulomb-like part of solution (3). Under 
this situation, a natural assumption will be that the quark interaction remains 
spherically symmetric at large distances too but then, if trying to extend 
the Coulomb-like part to large distances in a spherically symmetric way, we 
shall inevitably come to the solution (3) in virtue of the above theorem.  

Now one should say that the similar unique confining solutions exist for all 
semisimple and non-semisimple compact Lie groups, in particular, for SU($N$) 
with $N\ge2$ and 
U($N$) with $N\ge1$ \cite{{Gon051},{Gon052}}. Explicit form of solutions, 
e.g., for SU($N$) with $N=2,4$ can be found in Ref.\cite{Gon052} but it 
should be emphasized that components linear in $r$ always represent the 
magnetic (colour) field in all the mentioned solutions. Especially the case 
U(1)-group is interesting which corresponds to usual electrodynamics. 
Under this situation, as was pointed out in 
Refs. \cite{{Gon051},{Gon052}}, there is an interesting possibility of 
indirect experimental verification of the confinement mechanism under 
discussion. Indeed the confining solutions 
of Maxwell equations for classical electrodynamics point out 
the confinement phase could be in electrodynamics as well. Though 
there exist no elementary charged particles generating a constant magnetic 
field linear in $r$, the distance from particle, after all, if it could 
generate this elecromagnetic field configuration in laboratory then one might 
study motion of the charged particles in that field. The confining properties 
of the mentioned field should be displayed at classical level too but the exact 
behaviour of particles in this field requires certain analysis of the corresponding 
classical equations of motion. Such a program has been recently realized in 
Ref. \cite{GF10}. Motion of a charged (classical) particle was studied in the 
field representing magnetic part of the mentioned solution of Maxwell equations 
and it was shown that one deals with the full classical confinement of the 
charged particle in such a field: under any initial conditions the particle 
motion is accomplished within a finite region of space so that the particle 
trajectory is near magnetic field lines while the latter are compact manifolds 
(circles). Those results might be useful in thermonuclear plasma physics 
(for more details see \cite{GF10}). 

As has been repeatedly explained in 
Refs. \cite{{Gon051},{Gon052},{Gon031},{Gon032},{Gon04},{Gon06}}, 
parameters $A_{1,2}$ of 
solution (3) are inessential for physics in question and we can 
consider $A_1=A_2=0$. Obviously we have 
$\sum_{j=1}^{3}{\mathcal A}_{jt}=\sum_{j=1}^{3}{\mathcal A}_{j\varphi}=0$ which 
reflects the fact that for any matrix 
${\mathcal T}$ from SU(3)-Lie algebra we have ${\rm Tr}\,{\mathcal T}=0$. 
Also, as has been repeatedly discussed by us earlier (see, e. g., 
Refs. \cite{{Gon051},{Gon052}}), from the above form it is clear that 
the solution (3) is a configuration describing the electric Coulomb-like colour 
field (components $A^{3,8}_t$) and the magnetic colour field linear in $r$ 
(components $A^{3,8}_\varphi$) and we wrote down
the solution (3) in the combinations that are just 
needed further to insert into the Dirac equation (4). 

For the sake of completeness one should note that the similar unique confining 
solutions exist for all semisimple and 
non-semisimple compact Lie groups, in particular, for SU($N$) with $N\ge2$ and 
U($N$) with $N\ge1$ \cite{{Gon051},{Gon052}}. Explicit form of solutions, 
e.g., for SU($N$) with 
$N=2,4$ can be found in Ref. \cite{Gon052} but it should be emphasized that 
components linear in $r$ always represent the magnetic colour field in all the 
mentioned solutions. 

Another part of the family is given by the unique nonperturbative modulo 
square integrable solutions of the Dirac equation in the confining 
SU(3)-field of (3) $\Psi=(\Psi_1, \Psi_2, \Psi_3)$ 
with the four-dimensional Dirac spinors 
$\Psi_j$ representing the $j$th colour component of the meson, 
so $\Psi$ may describe the relative motion (relativistic bound states) of two 
quarks in mesons and the mentioned Dirac equation is written as follows 
$$i\partial_t\Psi\equiv  
i\pmatrix{\partial_t\Psi_1\cr \partial_t\Psi_2\cr \partial_t\Psi_3\cr}=
H\Psi\equiv$$
$$\pmatrix{H_1&0&0\cr 0&H_2&0\cr 0&0&H_3\cr}
\pmatrix{\Psi_1\cr\Psi_2\cr\Psi_3\cr}=
\pmatrix{H_1\Psi_1\cr H_2\Psi_2\cr H_3\Psi_3\cr}
                   \,,\eqno(4)$$
where the Hamiltonian $H_j$ is 
$$H_j=\gamma^0\biggl[\mu_0-i\gamma^1\partial_r-i\gamma^2\frac{1}{r}
\left(\partial_\vartheta+\frac{1}{2}\gamma^1\gamma^2\right)-$$
$$i\gamma^3\frac{1}{r\sin{\vartheta}}
\left(\partial_\varphi+\frac{1}{2}\sin{\vartheta}\gamma^1\gamma^3
+\frac{1}{2}\cos{\vartheta}\gamma^2\gamma^3\right)\biggr]-$$
$$g\gamma^0\left(\gamma^0{\mathcal A}_{jt}+\gamma^3\frac{1}{r\sin{\vartheta}}
{\mathcal A}_{j\varphi}\right) \eqno(5)  $$                           
with the gauge coupling constant $g$ while $\mu_0$ is a mass parameter and one 
should consider it to be the reduced mass which is equal to 
$ m_{q_1}m_{q_2}/(m_{q_1}+m_{q_2})$ with the current quark masses $m_{q_k}$ (
rest energies) of the corresponding quarks forming a meson (quarkonium). 

Then the unique nonperturbative modulo square integrable solutions of (4) 
are at $j=1,2,3$ (with Pauli matrix $\sigma_1$)  
$$\Psi_j=e^{-i\omega_j t}\psi_j\equiv 
e^{-i\omega_j t}r^{-1}\pmatrix{F_{j1}(r)\Phi_j(\vartheta,\varphi)\cr\
F_{j2}(r)\sigma_1\Phi_j(\vartheta,\varphi)}\>\eqno(6)$$
with the 2D eigenspinor $\Phi_j=\pmatrix{\Phi_{j1}\cr\Phi_{j2}}$ of the
Euclidean Dirac operator ${\mathcal D}_0$ on the unit sphere ${\mathbb S}^2$, while 
the coordinate $r$ stands for the distance between quarks. The explicit form of 
$\Phi_j$ is discussed in Appendix A. We can call the quantity 
$\omega_j$ the relative energy of the $j$th colour component of a meson (while $\psi_j$ is 
the wave function of a stationary state for the $j$th colour component), but 
we can see that if we want to interpret (4) as an equation for eigenvalues of 
the relative motion energy, i. e.,  to rewrite it in the form $H\psi=\omega\psi$ 
with $\psi=(\psi_1, \psi_2, \psi_3)$ then we should put $\omega=\omega_j$ for 
any $j$ so that $H_j\psi_j=\omega_j\psi_j=\omega\psi_j$. In this situation, 
if a meson is composed of quarks $q_{1,2}$ with different flavours then 
the energy spectrum of the meson will be given 
by $\epsilon=m_{q_1}+m_{q_2}+\omega$ with the current quark masses $m_{q_k}$ (
rest energies) of the corresponding quarks. On the other hand for 
determination of $\omega_j$ the following quadratic equation can be obtained 
\cite{{Gon01},{Gon051},{Gon052}}
$$[g^2a_j^2+(n_j+\alpha_j)^2]\omega_j^2-
2(\lambda_j-gB_j)g^2a_jb_j\,\omega_j+$$
$$[(\lambda_j-gB_j)^2-(n_j+\alpha_j)^2]g^2b_j^2-
\mu_0^2(n_j+\alpha_j)^2=0\>,  \eqno(7)   $$
which yields 
$$\omega_j=\omega_j(n_j,l_j,\lambda_j)=$$ 
$$\frac{\Lambda_j g^2a_jb_j\pm(n_j+\alpha_j)
\sqrt{(n_j^2+2n_j\alpha_j+\Lambda_j^2)\mu_0^2+g^2b_j^2(n_j^2+2n_j\alpha_j)}}
{n_j^2+2n_j\alpha_j+\Lambda_j^2}\>,$$
$$j=1,2,3\>,\eqno(8)$$

where $a_3=-(a_1+a_2)$, $b_3=-(b_1+b_2)$, $B_3=-(B_1+B_2)$, 
$\Lambda_j=\lambda_j-gB_j$, $\alpha_j=\sqrt{\Lambda_j^2-g^2a_j^2}$, 
$n_j=0,1,2,...$, while $\lambda_j=\pm(l_j+1)$ are
the eigenvalues of Euclidean Dirac operator ${\mathcal D}_0$ 
on a unit sphere with $l_j=0,1,2,...$. It should be noted that in the  
papers \cite{{Gon01},{Gon051},{Gon052},{Gon031},{Gon032},{Gon04},{Gon06}} 
we used the ansatz (6) 
with the factor $e^{i\omega_j t}$ instead of $e^{-i\omega_j t}$ but then the 
Dirac equation (4) would look as $-i\partial_t\Psi= H\Psi$ and in equation (7) 
the second summand would have the plus sign while the first summand in 
numerator of (8) would have the minus sign. In the papers 
\cite{{Gon07a},{Gon07b}} we returned to the conventional form of 
writing Dirac equation and this slightly modified the equations (7)--(8). In 
the given Chapter we conform to the same prescription as in 
Refs. \cite{{Gon07a},{Gon07b}}. 

In line with the above we should have $\omega=\omega_1=\omega_2=\omega_3$ in 
energy spectrum $\epsilon=m_{q_1}+m_{q_2}+\omega$ for any meson (quarkonium) 
and this at once imposes two conditions on parameters $a_j,b_j,B_j$ when 
choosing some experimental value for $\epsilon$ at the given current quark 
masses $m_{q_1},m_{q_2}$. 

The general form of the radial parts of (6) is considered in Appendix $B$. 
Within the given Chapter we need only the radial parts of (6) at $n_j=0$ 
(the ground state) that are [see $(B.5)$]  
$$F_{j1}=C_jP_jr^{\alpha_j}e^{-\beta_jr}\left(1-
\frac{gb_j}{\beta_j}\right), P_j=gb_j+\beta_j, $$
$$F_{j2}=iC_jQ_jr^{\alpha_j}e^{-\beta_jr}\left(1+
\frac{gb_j}{\beta_j}\right), Q_j=\mu_0-\omega_j\eqno(9)$$
with $\beta_j=\sqrt{\mu_0^2-\omega_j^2+g^2b_j^2}$, while $C_j$ is determined 
from the normalization condition
$\int_0^\infty(|F_{j1}|^2+|F_{j2}|^2)dr=\frac{1}{3}$. 
Consequently, we shall gain that $\Psi_j\in L_2^{4}({\mathbb R}^3)$ at any 
$t\in{\mathbb R}$ and, as a result,
the solutions of (6) may describe relativistic bound states (mesons) 
with the energy (mass) spectrum $\epsilon$.
\subsection{Nonrelativistic and the weak coupling limits}
It is useful to specify the nonrelativistic limit (when 
$c\to\infty$) for spectrum (8). For this one should replace 
$g\to g/\sqrt{\hbar c}$, 
$a_j\to a_j/\sqrt{\hbar c}$, $b_j\to b_j\sqrt{\hbar c}$, 
$B_j\to B_j/\sqrt{\hbar c}$ and, expanding (8) in $z=1/c$, we shall get
$$\omega_j(n_j,l_j,\lambda_j)=\pm\mu_0c^2\left[1\mp
\frac{g^2a_j^2}{2\hbar^2(n_j+|\lambda_j|)^2}z^2\right]$$
$$+\left[\frac{\lambda_j g^2a_jb_j}{\hbar(n_j+|\lambda_j|)^2}\,
\mp\mu_0\frac{g^3B_ja_j^2f(n_j,\lambda_j)}{\hbar^3(n_j+
|\lambda_j|)^{7}}\right]
z\,+O(z^2)\>,\eqno(10)$$
where 
$f(n_j,\lambda_j)=4\lambda_jn_j(n_j^2+\lambda_j^2)+
\frac{|\lambda_j|}{\lambda_j}\left(n_j^{4}+6n_j^2\lambda_j^2+\lambda_j^4
\right)$. 

As is seen from (10), at $c\to\infty$ the contribution of linear magnetic 
colour field (parameters $b_j, B_j$) to the spectrum really vanishes and the 
spectrum in essence becomes the purely nonrelativistic Coulomb one (modulo 
the rest energy). Also it is 
clear that when $n_j\to\infty$, $\omega_j\to\pm\sqrt{\mu_0^2+g^2b_j^2}$. 
At last, one should specify the weak 
coupling limit of (8), i.e., the case $g\to0$. As is not complicated to see 
from (8), $\omega_j\to\pm\mu_0$ when $g\to0$. But then quantities 
$\beta_j=\sqrt{\mu_0^2-\omega_j^2+g^2b_j^2}\to0$ and wave functions of (9) 
cease to be the modulo square integrable ones at $g=0$, i.e., they cease to 
describe relativistic bound states. Accordingly, this means that the equation 
(8) does not make physical meaning at $g=0$. 

We may seemingly use (8) with various combinations of signes ($\pm$) before 
the second summand in numerators of (8) but, due to (10), it is 
reasonable to take all signs equal to plus which is our choice within the 
Chapter. Besides, 
as is not complicated to see, radial parts in the nonrelativistic limit have 
the behaviour of form $F_{j1},F_{j2}\sim r^{l_j+1}$, which allows one to call 
quantum number $l_j$ angular momentum for the $j$th colour component though 
angular momentum is not conserved in the field (3) \cite{{Gon01},{Gon052}}. So, 
for mesons under consideration we should put all $l_j=0$. 
\subsection{Chiral limit}
There is one more interesting limit for relation (8) -- the chiral one, i.e., 
the situation when $m_{q1},m_{q2}\to0$ which entails $\mu_0\to0$ and (8) reduces 
to (at $j=1,2,3$)
$$(\omega_j)_{\rm chiral}=\frac{\Lambda_j g^2a_jb_j\pm(n_j+\alpha_j)g|b_j|
\sqrt{n_j^2+2n_j\alpha_j}}
{n_j^2+2n_j\alpha_j+\Lambda_j^2}\>,\eqno(8^\prime)$$
which mathematically signifies that Dirac equation in the field (3) 
possesses a nontrivial spectrum of bound states even for massless fermions. 
Physically this gives us  
a possible approach to the problem of chiral symmetry breaking in QCD 
\cite{Gon08b}: in chirally symmetric world masses of mesons are fully 
determined by the confining SU(3)-gluonic field between (massless) quarks 
and not equal to zero. Accordingly chiral symmetry is a sufficiently rough 
approximation holding true only when neglecting the mentioned SU(3)-gluonic 
field between quarks and no additional mechanism of the spontaneous chiral 
symmetry breaking connected to the so-called Goldstone bosons is required. 
As a result, e.g., masses of mesons from pseudoscalar nonet have a purely 
gluonic contribution and we have considered it in \cite{{Gon08b},{Gon10}} 
(see also sect. 6). 

One can note that for to be the nonzero chiral limit of (8) the crucial role 
belongs to the colour magnetic field linear in $r$ [parameters $b_{1,2}$ from 
solution (3)] inasmuch as chiral limit is equal exactly to zero when 
$b_{1,2}=0$. On the contrary, when parameters $a_{1,2}$ the Coulomb colour 
electric part of solution (3) are equal to zero the chiral limit may be nonzero 
at $b_{1,2}\ne0$, as is seen from $(8^\prime)$ except for the case $n_j=0$ when 
both parts of SU(3)-gluonic field (3) are important for confinement and mass 
generation in chiral limit.  

\subsection{Eigenspinors with $\lambda=\pm1$}
Finally it should be noted that spectrum (8) is degenerated owing to the  
degeneracy of eigenvalues for the
Euclidean Dirac operator ${\mathcal D}_0$ on the unit sphere ${\mathbb S}^2$. 
Namely, each eigenvlalue of ${\mathcal D}_0$ $\lambda =\pm(l+1), l=0,1,2...$, has 
multiplicity $2(l+1)$, so we has $2(l+1)$ eigenspinors orthogonal to each other. 
Ad referendum we need eigenspinors corresponding to $\lambda =\pm1$ ($l=0$) 
so here is their explicit form [see $(A.16)$]
$$\lambda=-1: \Phi=\frac{C}{2}\pmatrix{e^{i\frac{\vartheta}{2}}
\cr e^{-i\frac{\vartheta}{2}}\cr}e^{i\varphi/2},\>$$
or
$$\Phi=\frac{C}{2}\pmatrix{e^{i\frac{\vartheta}{2}}\cr
-e^{-i\frac{\vartheta}{2}}\cr}e^{-i\varphi/2},$$
$$\lambda=1: \Phi=\frac{C}{2}\pmatrix{e^{-i\frac{\vartheta}{2}}\cr
e^{i\frac{\vartheta}{2}}\cr}e^{i\varphi/2},$$
or
$$\Phi=\frac{C}{2}\pmatrix{-e^{-i\frac{\vartheta}{2}}\cr
e^{i\frac{\vartheta}{2}}\cr}e^{-i\varphi/2} 
\eqno(11) $$
with the coefficient $C=1/\sqrt{2\pi}$ (for more details, see 
Appendix $A$). 

\subsection{Meson energy (mass)}

Within the present Chapter we shall use relations (8) at $n_j=0=l_j$ so 
energy (mass) of a meson is given by $\mu=m_{q_1}+m_{q_2}+\omega$ 
with $\omega=\omega_j(0,0,\lambda_j)$ for any $j=1,2,3$ whereas 
$$\omega=\frac{g^2a_1b_1}{\Lambda_1}+\frac{\alpha_1\mu_0}
{|\Lambda_1|}=\frac{g^2a_2b_2}{\Lambda_2}+\frac{\alpha_2\mu_0}
{|\Lambda_2|}$$
$$=\frac{g^2a_3b_3}{\Lambda_3}+\frac{\alpha_3\mu_0}
{|\Lambda_3|}=\mu-m_{q_1}-m_{q_2}
\>\eqno(12)$$
and, as a consequence, the corresponding meson wave functions of 
(6) are represented by (9) and (11). 
\subsection{Choice of quark masses and the gauge coupling constant}
It is evident for employing the above relations we have to assign some values 
to quark masses and gauge coupling constant $g$. We take the current quark 
masses used in \cite{{Gon07a},{Gon07b},{Gon08b}} and they are
$m_u=2.25\>\,{\rm MeV}$, $m_d=5\>\,{\rm MeV}$, $m_s=107.5\>\,{\rm MeV}$. 
Under the circumstances, the reduced mass $\mu_0$ of (5) will be equal to 
$m_u/2$, $m_d/2$, $m_s/2$ accordingly. As to the 
gauge coupling constant $g=\sqrt{4\pi\alpha_s}$, it should be noted that 
recently some attempts have been made to generalize the standard formula
for $\alpha_s=\alpha_s(Q^2)=12\pi/[(33-2n_f)\ln{(Q^2/\Lambda^2)}]$ ($n_f$ is 
number of quark flavours) holding true at the momentum transfer 
$\sqrt{Q^2}\to\infty$ 
to the whole interval $0\le \sqrt{Q^2}\le\infty$. We shall employ one such a 
generalization used in Refs. \cite{De1}. It is written as follows 
($x=\sqrt{Q^2}$ in GeV) 
$$ \alpha(x)=\frac{12\pi}{(33-2n_f)}\frac{f_1(x)}{\ln{\frac{x^2+f_2(x)}
{\Lambda^2}}} 
\eqno(13) $$
with 
$$f_1(x)=$$
$$1+\left(\left(\frac{(1+x)(33-2n_f)}{12}\ln{\frac{m^2}{\Lambda^2}}-1
\right)^{-1}+0.6x^{1.3}\right)^{-1}\>,$$
$$f_2(x)=m^2(1+2.8x^2)^{-2}\>,$$
wherefrom one can conclude that $\alpha_s\to \pi=3.1415...$ when $x\to 0$, 
i. e., $g\to{2\pi}=6.2831...$. We used (13) at $m=1$ GeV, $\Lambda=0.234$ GeV, 
$n_f=3$, $x=1232.0 $ MeV or $x=1672.45 $ MeV to obtain $g\approx3.367244706$ 
or $g\approx2.841355055$ necessary for 
our further computations at the mass scale of $\Delta^{++}$, $\Delta^{-}$, 
or $\Omega^-$ respectively. 

\subsection{Electric form factor and the root-mean-square radius}
For each meson (quarkonium) with the wave function $\Psi=(\Psi_j)$ of (6) we 
can define the  
electromagnetic current $J^\mu=\overline{\Psi}(I_3\otimes\gamma^\mu)\Psi=
(\Psi^{\dag}\Psi,\Psi^{\dag}(I_3\otimes{\bf \alpha})\Psi)=(\rho,{\bf J})$, 
${\bf \alpha}=\gamma^0{\bf\gamma}$.  
The electric form factor $f(K)$ is the Fourier transform of $\rho$
$$ f(K)= \int\Psi^{\dag}\Psi e^{-i{\bf K}{\bf r}}d^3x=\sum\limits_{j=1}^3
\int\Psi_j^{\dag}\Psi_j e^{-i{\bf K}{\bf r}}d^3x =$$
$$\sum\limits_{j=1}^3f_j(K)=\sum\limits_{j=1}^3
\int (|F_{j1}|^2+|F_{j2}|^2)\Phi_j^{\dag}\Phi_j
\frac{e^{-i{\bf K}{\bf r}}}{r^2}d^3x,\>$$
$$d^3x=r^2\sin{\vartheta}dr d\vartheta d\varphi\eqno(14)$$
with the momentum transfer $K$. At $n_j=0=l_j$, as is easily seen, for any  
spinor of (11) we have $\Phi_j^{\dag}\Phi_j=1/(4\pi)$, so the integrand in 
(14) does not depend on $\varphi$ and we can consider vector ${\bf K}$ to be 
directed along z-axis. Then ${\bf Kr}=Kr\cos{\vartheta}$ and with the help of 
(9) and relations (see Ref. \cite{PBM1}): $\int_0^\infty 
r^{\alpha-1}e^{-pr}dr=
\Gamma(\alpha)p^{-\alpha}$, Re $\alpha,p >0$,
$\int_0^\infty r^{\alpha-1}e^{-pr}\pmatrix{\sin{(Kr)}\cr\cos{(Kr)}\cr}dr=$
$$\Gamma(\alpha)(K^2+p^2)^{-\alpha/2}
\pmatrix{\sin{(\alpha\arctan{(K/p))}}\cr\cos{(\alpha\arctan{(K/p))}}\cr},$$ 
Re $\alpha >-1$, Re $p > |{\rm Im}\, K|$, $\Gamma(\alpha+1)=\alpha\Gamma(\alpha)$, 
$$\int_0^\pi e^{-iKr\cos{\vartheta}}\sin{\vartheta}d\vartheta=
2\sin{(Kr)}/(Kr),$$ 
we shall obtain 
$$ f(K)=\sum\limits_{j=1}^3f_j(K)=$$
$$\sum\limits_{j=1}^3\frac{(2\beta_j)^{2\alpha_j+1}}{6\alpha_j}\cdot
\frac{\sin{[2\alpha_j\arctan{(K/(2\beta_j))]}}}{K(K^2+4\beta_j^2)^{\alpha_j}}$$
$$=\sum\limits_{j=1}^3\left(\frac{1}{3}-\frac{2\alpha^2_j+3\alpha_j+1}
{6\beta_j^2}\cdot \frac{K^2}{6}\right)+O(K^4), \eqno(15)$$
wherefrom it is clear that $f(K)$ is a function of $K^2$, as should be, and 
we can determine the root-mean-square radius of meson (quarkonium) in the form 
$$<r>=\sqrt{\sum\limits_{j=1}^3\frac{2\alpha^2_j+3\alpha_j+1}
{6\beta_j^2}}.\eqno(16)$$
When calculating (15) also the fact was used that by virtue of the 
normalization condition for wave 
functions we have $C_j^2[P_j^2(1-gb_j/\beta_j)^2+Q_j^2(1+gb_j/\beta_j)^2]=
(2\beta_j)^{2\alpha_j+1}/[3\Gamma(2\alpha_j+1)]$.

It is clear, we can directly calculate $<r>$ in accordance with the standard 
quantum mechanics rules as $<r>=\sqrt{\int r^2\Psi^{\dag}\Psi d^3x}=
\sqrt{\sum\limits_{j=1}^3\int r^2\Psi^{\dag}_j\Psi_j d^3x}$ and the 
result will be the same as in (16). So we should not call $<r>$ of (16) 
the {\em charge} radius of meson (quarkonium)-- it is just the radius of meson 
(quarkonium) determined 
by the wave functions of (6) (at $n_j=0=l_j$) with respect to strong 
interaction, i.e., radius of confinement. 
Now we should note the expression (15) to depend on 3-vector ${\bf K}$. To 
rewrite it in the form holding true for any 4-vector $Q$, let us recall that 
according to general considerations (see, e.g., Ref. \cite{LL1}) the relation 
(15) should correspond to the so-called Breit frame where $Q^2=-K^2$ 
[when fixing the metric by (1)] so it is 
not complicated to rewrite (15) for arbitrary $Q$ in the form 
$$ f(Q^2)=\sum\limits_{j=1}^3f_j(Q^2)=$$
$$\sum\limits_{j=1}^3\frac{(2\beta_j)^{2\alpha_j+1}}{6\alpha_j}\cdot
\frac{\sin{[2\alpha_j\arctan{(\sqrt{|Q^2|}/(2\beta_j))]}}}
{\sqrt{|Q^2|}(4\beta_j^2-Q^2)^{\alpha_j}}\> \eqno(17) $$
which passes on to (15) in the Breit frame. 

\subsection{Magnetic moment}
We can define the volumetric magnetic moment density by 
${\bf m}=q({\bf r}\times {\bf J})/2=q[(yJ_z-zJ_y){\bf i}+
(zJ_x-xJ_z){\bf j}+(xJ_y-yJ_x){\bf k}]/2$ with the meson charge $q$ and 
${\bf J}=\Psi^{\dag}(I_3\otimes{\bf \alpha})\Psi$. Using (6) we have in the 
explicit form 
$$J_x=\sum\limits_{j=1}^3
(F^\ast_{j1}F_{j2}+F^\ast_{j2}F_{j1})\frac{\Phi_j^{\dag}\Phi_j}
{r^2},\>$$ 
$$J_y=\sum\limits_{j=1}^3
(F^\ast_{j1}F_{j2}-F^\ast_{j2}F_{j1})
\frac{\Phi_j^{\dag}\sigma_2\sigma_1\Phi_j}{r^2},\>$$
$$J_z=\sum\limits_{j=1}^3
(F^\ast_{j1}F_{j2}-F^\ast_{j2}F_{j1})
\frac{\Phi_j^{\dag}\sigma_3\sigma_1\Phi_j}{r^2}  \eqno(18)$$
with Pauli matrices $\sigma_{1,2,3}$.
Magnetic moment of meson (quarkonium) is ${\bf M}=\int_V {\bf m}d^3x$, where 
$V$ is the volume 
of the meson (quarkonium) (the ball of radius $<r>$). Then at $n_j=l_j=0$, 
as is seen from (9), (11), $F^\ast_{j1}=F_{j1},F^\ast_{j2}=-F_{j2}$, 
$\Phi_j^{\dag}\sigma_2\sigma_1\Phi_j=0 $ for any spinor of (11) which entails 
$J_x=J_y=0$, i.e., $m_z=0$ while $\int_V m_{x,y}d^3x=0$ because of 
turning the integral over $\varphi$ to zero, which is easy to check.
As a result, the magnetic moments of mesons (quarkonia) with the 
wave functions of (6) (at $l_j=0$) are equal to zero, as should be according 
to experimental data \cite{pdg}. 
\subsection{Magnetic form factor and anomalous magnetic moment}
Following Refs. \cite{{Gon07a},{Gon07b}} we can, however, define magnetic form 
factor $F(K)$ of a meson if noticing that 
$$\int_V\sqrt{m_x^2+m_y^2+m_z^2}d^3x\ne$$
$$\sqrt{\left(\int_Vm_xd^3x\right)^2+
\left(\int_Vm_yd^3x\right)^2+\left(\int_Vm_zd^3x\right)^2}$$
$$=M=0\>.$$   
Under the circumstances we should define $F(K)$ as the inverse Fourier 
transform of $\sqrt{m_x^2+m_y^2+m_z^2}$
$$ \sqrt{m_x^2+m_y^2+m_z^2}=
\frac{q}{(2\pi)^3}\int F(K)e^{i{\bf K}{\bf r}}d^3K\>,$$
so that 
$$F(K)=
\frac{1}{q}\int \sqrt{m_x^2+m_y^2+m_z^2}e^{-i{\bf K}{\bf r}}d^3x=$$
$$\frac{|q|}{q}\int r|J_z\sin{\vartheta}|e^{-i{\bf K}{\bf r}}d^3x=$$
$$\frac{1}{2\pi}\frac{|q|}{q}\int\sum_{j=1}^3|F_{j1}||F_{j2}|
\frac{\sin^2{\vartheta}}{r}e^{-i{\bf K}{\bf r}}d^3x\>\eqno(19)$$
for any spinor of (11) so the integrand in (19) does not again depend on 
$\varphi$. Then ${\bf Kr}=Kr\cos{\vartheta}$ and 
using the relation 
$$\int_0^\pi e^{-iKr\cos{\vartheta}}\sin^3{\vartheta}d\vartheta=$$
$$4[\sin{(Kr)}/(Kr)^3-\cos{(Kr)}/(Kr)^2],$$ 
we shall obtain                             
$$F(K)=$$
$$2\frac{|q|}{q}\int_0^\infty r\sum_{j=1}^3C_j^2P_jQ_jr^{2\alpha_j}
e^{-2\beta_jr}
\left(1-\frac{g^2b_j^2}{\beta_j^2}\right)\times$$
$$\left[\frac{\sin{(Kr)}}{(Kr)^3}-\frac{\cos{(Kr)}}{(Kr)^2}\right]dr$$
$$=\frac{1}{3}\frac{|q|}{q}\sum_{j=1}^{3}\frac{(2\beta_j)^{2\alpha_j+1}}
{\alpha_j}\cdot\frac{{\mathcal P}_j{\mathcal Q}_j}
{{\mathcal P}_j^2+{\mathcal Q}_j^2}\cdot
\frac{1}{K^2(K^2+4\beta_j^2)^{\alpha_j}}\cdot$$
$$\left\{\frac{\sin{[(2\alpha_j-1)\arctan{(K/(2\beta_j))]}}}
{(2\alpha_j-1)K(K^2+4\beta_j^2)^{-1/2}}-
\cos{[2\alpha_j\arctan{(K/(2\beta_j))]}}  \right\}$$
$$=\frac{2}{3}\frac{|q|}{q}\sum_{j=1}^{3}\frac{\beta_j}{\alpha_j}
\cdot\frac{{\mathcal P}_j{\mathcal Q}_j}
{{\mathcal P}_j^2+{\mathcal Q}_j^2}
\biggl[\frac{\alpha_j(2\alpha_j+1)}{6\beta_j^2}-$$
$$\frac{\alpha_j(4\alpha_j^3+12\alpha_j^2+11\alpha_j+3)}{120\beta_j^4}K^2+
O(K^4)\biggr]$$
$$=\frac{2}{3}\frac{|q|}{q}\sum_{j=1}^{3}\frac{2\alpha_j+1}{6\beta_j}
\cdot\frac{{\mathcal P}_j{\mathcal Q}_j}
{{\mathcal P}_j^2+{\mathcal Q}_j^2}
\left(1-
3\frac{2\alpha_j^2+5\alpha_j+3}{10\beta_j^2}\cdot\frac{K^2}{6}\right)+
O(K^4)
\eqno(20)$$
with ${\mathcal P}_j=P_j(1-{gb_j}/{\beta_j})$, 
${\mathcal Q}_j=O_j(1+{gb_j}/{\beta_j})$. 
It is clear from (20) that 
$F(K)$ is a function of $K^2$ and we can rewrite (20) for arbitrary 4-vector 
$Q$ by the same manner as was done for electric form factor $f(K)$ in (17). 

Now we can define {\em the anomalous magnetic moment}\ for a meson by the 
relation 
$$\mu_a= F(K=0)=\frac{2}{3}\frac{|q|}{q}\sum_{j=1}^{3}\frac{2\alpha_j+1}{6\beta_j}
\cdot\frac{{\mathcal P}_j{\mathcal Q}_j}
{{\mathcal P}_j^2+{\mathcal Q}_j^2}\>,  \eqno(21)$$
as follows from (20). 
It should be noted that a possibility of existence 
of anomalous magnetic moments for mesons is permanently discussed in literature 
(see, e.g., Refs. \cite{{Maga},{Magb},{Magc}} and references therein) though there is no 
experimental evidence in this direction \cite{pdg}. Another matter is baryons 
with their nonzero anomalous magnetic moments and the latter are some 
essential characteristics of baryons. So that the relation (21) will be 
very useful to us below.
\subsection{Remarks about the relativistic two-body problem}
It is well known (see any textbook on quantum mechanics, e.g., Ref. \cite{Lev71}) 
that {\em nonrelativistic} two-body problem, when the particles interact with 
potential $V({\bf r_1},{\bf r_2})$ 
depending only on $|{\bf r_2-r_1}|=r$, reduces to the motion of one particle 
with reduced mass $m=m_1m_2/(m_1+m_2)$ in potential $V(r)$, where $r$ becomes  
the ordinary spherical coordinate from triplet $(r,\vartheta,\varphi)$. I.e., the 
bound states of {\em two} particles are the bound states of the particle with 
mass $m$ in potential $V(r)$ and they are the modulo square integrable 
solutions of the corresponding Schr{\"o}dinger equation. Another matter is 
{\em relativistic} two-body problem. As we emphasized in 
Refs. \cite{{Gon051},{Gon052}}, up to now it has no single-valued statement. But if 
noting that the 
most fundamental results of nonrelativistic quantum mechanics (hydrogen atom and 
so on) are connected with potentials $V(r)$ which are a part of the 
{\em electromagnetic} field $A=(A_0,{\bf A})$, i.e. $V(r)=A_0,{\bf A}=0$ then 
one may propose some formulation of the relativistic two-body problem. Really, 
now $V(r)$ additionally obeys the Maxwell equations and if we want to generalize 
the corresponding Schr{\"o}dinger equation to include an interaction with 
arbitrary electromagnetic field for ${\bf A}\ne 0$ then the answer is known: 
this is the Dirac equation with the replacement 
$\partial_\mu\to \partial_\mu-igA_\mu$, $g$ is a gauge coupling constant. 
Indeed, when going back in nonrelativistic limit at the light velocity 
$c\to\infty$ the magnetic field ${\bf A}$ vanishes because, as is well known, 
in the world with $c=\infty$ there exist no magnetic fields 
(see any elementary textbook on physics). At the same time $V(r)=A_0$ does 
not vanish and remains the same as in nonrelativistic case and the 
Dirac equation turns into the Schr{\"o}dinger equation (see Ref. \cite{Lev71}). 
But then we can see that $m$ in the above Dirac equation should consider 
the same reduced mass as before since in nonrelativistic limit we again should 
come to the standard formulation of two-body problem through effective particle 
with reduced mass $m$. So we can draw the conclusion that if
an electromagnetic field is a combination of electric $V(r)=A_0$ field 
between two charged elementary particles and some magnetic field 
${\bf A}={\bf A}(r)$ (which may be generated 
by the particles themselves and also depends only on $r$, distance between 
particles) then there are certain grounds to consider the 
given (quantum) relativistic two-body problem to be equivalent to the one of 
motion for one particle with usual reduced mass in the mentioned 
electromagnetic field. As a result, we can use the Dirac equation for finding 
possible relativistic bound states for such a particle implying that this is 
really some description of the corresponding two-body problem. Under this 
situation we should remark the following.

1. Although for simplicity we talked about electromagnetic field but everything 
holds true for any Yang-Mills field, in particular, for SU(3)-gluonic field 
while the Maxwell equations are replaced by the Yang-Mills ones.

2. There arises the question: whether the Maxwell or Yang-Mills equations possess 
solutions with spherically symmetric $A_0(r), {\bf A}(r)$? The answer is 
given by the uniqueness theorem (see subsection 2.1 and Appendix D). 

3. The Dirac equation in such a field has the nonperturbative spectrum (8)  
and the latter should be treated as the nonpeturbative {\em interaction 
energy} of two quarks and $r$ as the distance between quarks and so on 
(see Section 2). So eq. (4) should be understood just in this 
manner. In line with the above we should have $\omega=\omega_1=\omega_2=
\omega_3$ (with $\omega_j$ of (8)) in 
energy spectrum $\epsilon=m_{q_1}+m_{q_2}+\omega$ for any meson (quarkonium) 
and this at once imposes two conditions on parameters $a_j,b_j,B_j$ of 
solution (3) when choosing some experimental value for $\epsilon$ at the given 
current quark masses $m_{q_1},m_{q_2}$. At last, the Dirac equation in question 
is obtained from QCD lagrangian (with one flavor) if mass parameter in the 
latter is taken to be equal to the above reduced mass. Therefore, we can say 
that meson wave functions (6) are the nonperturbative solutions of 
the Dirac-Yang-Mills system directly derived from QCD-lagrangian.

4. To summarize, there exist good physical and mathematical grounds for 
formulation of the above relativistic two-body problem that has a correct 
nonrelativistic limit and is Lorentz and gauge invariant. It is clear that all 
the above considerations can be justified only by comparison with experimental 
data but now we obtain some intelligible 
programme of further activity which has been partly realized in 
many our papers cited above. 

Much of the above was disscussed in Refs. \cite{{Gon051},{Gon052}} but perhaps 
in other words. 

\section{Wave functions of baryons}
\subsection{Preliminaries}
In accordance with SQM \cite{pdg} baryons are composed from three quarks, 
generally speaking, with different flavours, so a baryon can be represented in 
the schematic form shown in Fig. 1, where $r_i$ are distances between 
the corresponding quarks. 
\begin{figure}
\vspace{0cm}
\caption{Schematic sketch of a baryon according to SQM.}
\end{figure}
As a consequence, we are faced with a relativistic 3-body problem. Under the 
circumstances the main task is to obtain an interaction Hamiltonian $H$ of the 
3-quark system if taking that every two quarks interact through the gluonic field 
described by solution (3), in general, with different constants $a_j$, 
$b_j$, $B_j$ for each pair of quarks. Then energy levels of a baryon will be 
given by relation $\epsilon=m_{q_1}+m_{q_2}+m_{q_3}+\omega$ with the current 
quark masses $m_{q_k}$ (rest energies) of the quarks constituting the baryon 
while the interaction energy $\omega$ should be subject to equation
$H\psi=\omega\psi$ with a baryonic wave function $\psi$ describing relative 
motion (bound state) of three quarks.

To construct wave functions of baryons let us at first make a remark concerning 
the picture of interaction between two quarks which was obtained when applying 
the confinement mechanism in question to description of heavy quarkonia and 
mesons from pseudoscalar nonet 
\cite{{Gon031},{Gon032},{Gon08a},{Gon06},{Gon07a},{Gon07b},{Gon08},{Gon08b},{Gon10}}. 
\subsection{Qualitative picture of quark interaction}
Let us recall that, according to Refs. \cite{{Gon052},{Gon06}}, one can 
confront the field (3) with the $T_{00}$-component (the volumetric energy 
density of the SU(3)-gluonic field) of the energy-momentum tensor (2) so that 
$$T_{00}\equiv T_{tt}=\frac{E^2+H^2}{2}=$$
$$\frac{1}{2}\left(\frac{a_1^2+
a_1a_2+a_2^2}{r^4}+\frac{b_1^2+b_1b_2+b_2^2}{r^2\sin^2{\vartheta}}\right)
\equiv\frac{{\mathcal A}}{r^4}+
\frac{{\mathcal B}}{r^2\sin^2{\vartheta}}\>\eqno(22)$$
with electric $E$ and magnetic $H$ colour field strengths and with 
real ${\mathcal A}>0$, ${\mathcal B}>0$. One can also introduce magnetic colour 
induction $B=(4\pi\times10^{-7} {\rm H/m})\,H$, where $H$ in A/m.  

We should now note that one can calculate energy ${\mathcal E}$ of gluon 
condensate conforming to solution (3) in a volume $V$ through relation 
${\mathcal E}=\int_VT_{00}r^2\sin{\vartheta}dr d\vartheta d\varphi\>$ with 
$T_{00}$ of (22) but one should take into account that classical $T_{00}$ has a 
singularity along $z$-axis ($\vartheta=0,\pi$) and we have to introduce some 
angle $\vartheta_0$ so $\vartheta_0\leq\vartheta\leq\pi-\vartheta_0$.  
As well as in Refs. \cite{{Gon051},{Gon06},{Gon08b}}, we may consider 
$\vartheta_0$ to be a parameter determining some cone $\vartheta=\vartheta_0$ 
so the quark emits gluons outside of the cone. Now if there are two quarks 
$Q_1, Q_2$ and each of them emits gluons outside 
of its own cone $\vartheta=\vartheta_{1,2}$ (see Figs. 2, 3) then we have 
soft gluons (as mentioned in section 1) in regions I, II and between quarks.  

Accordingly, we shall have some region $V$ with gluon condensate between quarks 
$Q_1, Q_2$ and its vertical projection is shown in Fig. 2. Another projection 
of $V$ onto a plane perpendicular to the one of Fig. 2 is sketched out in 
Fig. 3.  
\begin{figure}
\vspace{0cm}
\caption{Vertical projection of region with the gluon condensate energy 
between quarks.}
\end{figure}

\begin{figure}
\vspace{0cm}
\caption{Horizontal projection of region with the gluon condensate energy 
between quarks.}
\end{figure}

To clarify a physical meaning of the quantities $r_{1,2}$ in Figs. 2, 3, 
let us recall an analogy with classical 
electrodynamics where is well known (see, e. g., \cite{LL}) that the notion 
of classical electromagnetic field (a photon condensate) generated by a 
charged particle is applicable only at distances much greater than the Compton 
wavelength 
$\lambda_c=1/m$ for the given particle with mass $m$. Within the QCD framework 
the parameter $\Lambda_{QCD}$ plays a similar part (see, e.g., 
Ref. \cite{pdg}). 
Namely, the notion of classical SU(3)-gluonic field ( a gluon condensate) is 
not applicable at the distances much 
less than $1/\Lambda_{QCD}$. In accordance with subsection 2.6 we took 
$\Lambda_{QCD}=\Lambda=0.234$ GeV which entails $1/\Lambda\sim$ 0.8433 fm 
so one may consider $r_{1,2}\sim0.08$ fm. 

If taking into account that only the values of $\vartheta_{1,2}$, 
$\varphi_{1,2}$ between 0 and $90^\circ$ are of physical meaning then it 
should be noted that
the estimates of $\vartheta_{1,2}$, $\varphi_{1,2}$ obtained in 
Refs. \cite{{Gon06},{Gon08b}} show that $\vartheta_{1,2}\to\pi/2$, 
$\varphi_{1,2}\to0$ with increasing the quark masses so gluon condensate
(a classical SU(3)-gluonic field) is mainly concentrated near the axis 
connecting quarks. Under the circumstances we may represent a baryon composed 
from three pairs of the interacting quarks in the schematic form of Fig. 4, 
where the shaded areas schematically stand for soft gluons between quarks. 
\begin{figure}
\vspace{0cm}
\caption{Schematic sketch of quark interactions in a baryon.}
\end{figure}
As a consequence, we may suppose main interaction of quarks in a baryon to be 
concentrated along the axes joining quarks in each pair while one may 
practically neglect possible interaction of any pair of quarks as a whole with 
third remaining quark. When approaching all three quarks the latter interaction 
should not increase due to asymptotical freedom so the given picture may 
in fact hold true at the scales of baryon. In what follows we consider that is 
the case and then we can construct the baryonic wave functions in the next 
way.
\subsection{Baryonic wave functions and the interaction Hamiltonian}
In view of the above it is natural to characterize interaction of each pair of 
quarks in a baryon by the Hamiltonian of form (4)--(5) and, accordingly, 
relative motion of quarks in the pair by spherical coordinates $r_i$, 
$\vartheta_i$, $\varphi_i$, 
where $i=1$, $2$, $3$ correspond to pairs $q_1q_2$, $q_1q_3$, $q_2q_3$ 
respectively (see Fig. 1, 4). Then, in accordance with (4)--(5) we can 
introduce the Hamiltonians $H^{(i)}=(H^{(i)}_j)$ with colour index $j=1,2,3$.  
In their turn, $H^{(i)}$ will depend on $a_j^{(i)}$, $b_j^{(i)}$, $B_j^{(i)}$, 
i.e., the conforming parameters of solution (3) describing interaction of the 
$i$th pair of quarks [cf. (5)] while $\mu_0$ will be the reduced mass 
of the corresponding quark pair. Each $H^{(i)}$ acts in Hilbert space 
$L_2^{12}({\mathbb R}^3)$ which possesses the basis consisting of the 
eigenfunctions of $H^{(i)}$ and having the form of (6) (with the corresponding 
change of notations and omitting the phase factor $e^{-i\omega_j t}$). In 
what follows we denote such a basis by $\psi^{(i)}=(\psi^{(i)}_j)$ according 
to (6). In virtue of smallness of baryon sizes we may put the time variables 
$t_i$ for each quark pair to be equal, i.e., $t_1=t_2=t_3$ and then the 
stationary states of a baryon can be chosen in the form of tensorial product 
of $\psi^{(i)}$ (see Appendix C)
$$\psi=\psi^{(1)}\otimes\psi^{(2)}\otimes\psi^{(3)}\>,\eqno(23)$$
so $\psi$ will be a basis in the Hilbert space 
${\mathcal H}=L_2^{12}({\mathbb R}^3)\otimes L_2^{12}({\mathbb R}^3)
\otimes L_2^{12}({\mathbb R}^3)$ and 
also the eigenfunctions for operator in ${\mathcal H}$ 
(with identical operator $I$)  
$$ H=H^{(1)}\otimes{I}\otimes{I}+{I}\otimes H^{(2)}\otimes{I}+
{I}\otimes{I}\otimes H^{(3)}     \eqno(24)$$
with the eigenvalues $\omega=\omega_j^{(1)}+\omega_j^{(2)}+\omega_j^{(3)}$, 
where $\omega_j^{(i)}$ is an eigenvalue for operator $H^{(i)}_j$ (see 
subsection 2.1 and Appendix C), i.e., 
$$H\psi=H(\psi^{(1)}\otimes\psi^{(2)}\otimes\psi^{(3)})=\omega\psi=
(\omega_j^{(1)}+\omega_j^{(2)}+\omega_j^{(3)})\psi\>.   \eqno(25)$$
It is clear that operator $H$ is just the sought interaction Hamiltonian 
for quarks in a baryon. As a result, the energy levels of the baryon will be 
given by relation $\epsilon=m_{q_1}+m_{q_2}+m_{q_3}+\omega$ with the current 
quark masses $m_{q_k}$ (rest energies) of the quarks constituting the baryon 
and the interaction energy $\omega$ of (25).  

Under the situation, the stay probability of a baryon in the volume element 
$dV_1dV_2dV_3$ of the configuration space with 
$dV_i=r_i^2\sin{\vartheta_i}dr_id\vartheta_id\varphi_i$ will be equal to
$$dP=(\psi^{+{(1)}}\psi^{(1)})(\psi^{+{(2)}}\psi^{(2)})(\psi^{+{(3)}}\psi^{(3)})
dV_1dV_2dV_3  \eqno(26)$$
in accordance with (C.3), so 
$$\int_{{\mathbb R}^3}\int_{{\mathbb R}^3}\int_{{\mathbb R}^3} 
dP\,=1.\eqno(27)$$
Having integrated $dP$ over the angles $\vartheta_i,\varphi_i$ (with taking into 
account the properties of the eigenspinors of the Euclidean Dirac operator on 
two-sphere ${\mathbb S}^2$, see Appendix A), we obtain the probability of that 
the quarks in the baryon are at the distances $r_i$ from each other (see Fig. 4)
$$dW=\prod_{i=1}^{3}\left(\sum_{j=1}^{3}(|F_{j1}^{(i)}|^2+
|F_{j2}^{(i)}|^2)dr_i\right). 
\eqno(28)$$

\section{Main characteristics of symmetrical baryons}
Obviously, we should choose a few quantities that are the most important from 
the physical point of view to characterize 
baryons under consideration and then we should evaluate the given quantities 
within the framework of our approach. In the circumstances let us settle on 
the ground state energy (mass), the root-mean-square 
radius and the anomalous magnetic moment. All three magnitudes are essentially 
nonperturbative ones, and can be calculated only by nonperturbative techniques.
\subsection{Ground state energy (mass)}
One should evidently take into account that for symmetrical baryons in question 
from physical point of view we have all the grounds of considering the 
interaction in pairs $q_1q_2$, $q_1q_3$, $q_2q_3$ (see Figs. 1, 4) to be the 
same because $q_1=q_2=q_3$ and, as a result, the corresponding parameters 
$a_j^{(i)}$, $b_j^{(i)}$ $B_j^{(i)}$ of gluonic field from solution (3) 
describing interaction of the $i$th pair of quarks should be equal: 
$a_j^{(i)}=a_j$, $b_j^{(i)}=b_j$, $B_j^{(i)}=B_j$. Then, in accordance with 
(12) and (25) we obtain the symmetrical baryon mass as 
$$\mu=3(m_q+\omega),\, \omega=\omega_j=
\frac{g^2a_jb_j}{\Lambda_j}+\frac{\alpha_j\mu_0}
{|\Lambda_j|}\,, \eqno(29)$$
or, in more explicit form, 

$$\omega=\frac{g^2a_1b_1}{\lambda_1-gB_1}+
\mu_0\frac{\sqrt{(\lambda_1-gB_1)^2-g^2a_1^2}}
{|\lambda_1-gB_1|}=$$ 
$$\frac{g^2a_2b_2}{\lambda_2-gB_2}+
\mu_0\frac{\sqrt{(\lambda_2-gB_2)^2-g^2a_2^2}}
{|\lambda_2-gB_2|}=$$
$$\frac{g^2(a_1+a_2)(b_1+b_2)}{\lambda_3+g(B_1+B_2)}+
\mu_0\frac{\sqrt{[\lambda_3+g(B_1+B_2)]^2-g^2(a_1+a_2)^2}}{|\lambda_3+
g(B_1+B_2)|}\>\eqno(30)$$
with the reduced mass $\mu_0=m_q/2$ while in what follows we take the 
eigenvalues of the Euclidean Dirac operator on ${\mathbb S}^2$ 
$\lambda_j=1$, $j=1,2,3$. 

\subsection{Baryon radius}
We can note that the root-mean-square radius of meson from (16) can be defined 
by the relation 
$$<r>^2=-6\frac{\partial^2F}{\partial K^2}(K=0) \eqno(31)$$
with form factor $F(K)$ of (15). Obviously for symmetrical baryons 
it is natural to put $<r_1>^2=<r_2>^2=<r_3>^2=r_0^2$, where $<r_i>^2$ conforms 
to the corresponding quark pair (see Figs. 1, 4). But then simple geometric 
considerations give rise 
to the baryon radius $<R>={r_0}/\sqrt{3}$ which can be defined by the relation 
analogous to (31)
$$<R>^2=-6\frac{\partial^2F_E}{\partial K^2}(K=0) \eqno(32)$$
if putting the baryon electric form factor 
$F_E(K)=\frac{1}{9}\sum\limits_{j=1}^3\,F_i(K)$, where form factor $F_i(K)$ 
corresponds to $i$th quark pair (see Figs. 1, 4), which entails the expression 
$$<R>=\frac{1}{\sqrt{3}}\sqrt{\sum\limits_{j=1}^3\frac{2\alpha^2_j+3\alpha_j+1}
{6\beta_j^2}}\eqno(33)$$
for symmetrical baryons. 

\subsection{Anomalous magnetic moment}
Also it is natural to define the baryon magnetic form factor by the relation 
$F_M(K)=\frac{1}{3}\sum\limits_{j=1}^3\,F_i(K)$, where form factor $F_i(K)$ 
of (20) corresponds to $i$th quark pair (see Figs. 1, 4), which gives rise to 
the expression for the baryon anomalous magnetic moment         

$$\mu_a= F_M(K=0)=\frac{2}{3}\frac{|q|}{q}\sum_{j=1}^{3}\frac{2\alpha_j+1}{6\beta_j}
\cdot\frac{{\mathcal P}_j{\mathcal Q}_j}
{{\mathcal P}_j^2+{\mathcal Q}_j^2}\,,$$
$${\mathcal P}_j=P_j(1-{gb_j}/{\beta_j})\,,
{\mathcal Q}_j=O_j(1+{gb_j}/{\beta_j})
  \eqno(34)$$
in the case of symmetrical baryons 
while $q$ is electric charge of $i$th quark pair.

\section{Numerical estimates}
\subsection{$\Delta^{++}$}
At computation we took mass $\mu=1232.0$ MeV \cite{pdg} while the anomalous 
magnetic moment is experimentally restricted within limits $3.7\le\mu_a\le7.5$ 
\cite{pdg} in units of nuclear magneton $\mu_N$, so under calculation we took 
$\mu_a=(3.7+7.5)/2=5.6$. As for the radius $<R>$, no experimental values 
of it exist \cite{pdg} but if taking into account that in the case of proton with 
mass 938 MeV we have $<R>\approx0.875$ fm \cite{pdg}, then when considering 
the symmetrical baryons to be more compact than proton we should put, for 
example, 0.3 fm $\le<R>\le$ 0.7 fm to make computation more sensible. Some 
numerical results are adduced in Tables 1 and 2 and each line of Table 2 
corresponds to the conforming one of Table 1 while the values of $\mu_a$ 
are expressed in units of $\mu_N$. 

\begin{table*}
\caption{Gauge coupling constant, reduced mass $\mu_0$ and
parameters of the confining SU(3)-gluonic field for $\Delta^{++}$}
\label{t.1}
\begin{center}
\begin{tabular}{|c|c|c|c|c|c|c|c|}
\hline
\small $ g$ & \small $\mu_0$ (\small MeV) & \small $a_1$ 
& \small $a_2$ & \small $b_1$ (\small GeV) & \small $b_2$ (\small GeV) 
& \small $B_1$ & \small $B_2$ \\
\hline
\scriptsize 3.36724
& \scriptsize 1.12500
& \scriptsize 0.550635
& \scriptsize -0.764655
& \scriptsize -0.175434
& \scriptsize -0.170501
& \scriptsize  1.09500
& \scriptsize   -0.7800 \\
\hline
\scriptsize 3.36724
& \scriptsize 1.12500
& \scriptsize 0.404093
& \scriptsize 0.203740
& \scriptsize -0.306289
& \scriptsize 0.514750
& \scriptsize 1.3200
& \scriptsize -0.5700 \\
\hline
\scriptsize 3.36724
& \scriptsize 1.12500
& \scriptsize 0.432052
& \scriptsize -0.136707
& \scriptsize 0.318374
& \scriptsize -0.780394
& \scriptsize -0.8400
& \scriptsize -0.58500 \\
\hline
\end{tabular}
\end{center}
\end{table*}
\newpage
\begin{table*}
\caption{Theoretical and experimental mass, radius and anomalous magnetic 
moment for $\Delta^{++}$}
\label{t.2}
\begin{center}
\begin{tabular}{|c|c|c|c|c|c|} 
\hline
\tiny Theoret. $\mu$ (MeV) 
&  \tiny Experim. $\mu$ (MeV) 
& \tiny Theoret. $<R>$ (fm)  
& \tiny Experim. $<R>$ (fm) 
& \tiny Theoret. $\mu_a$  
& \tiny Experim. $\mu_a$  \\
\hline
\scriptsize 1232.0
& \scriptsize $\mu= 3(m_u+\omega_j(0,0,1))= 1232.0$
& \scriptsize 0.696633
& \scriptsize -
& \scriptsize 5.37931
& \scriptsize 5.6 \\
\hline
\scriptsize 1232.0
& \scriptsize $\mu= 3(m_u+\omega_j(0,0,1))= 1232.0$
& \scriptsize 0.516086
& \scriptsize -
& \scriptsize 5.77838
& \scriptsize 5.6 \\
\hline
& \scriptsize $\mu= 3(m_u+\omega_j(0,0,1))= 1232.0$
& \scriptsize 0.356102
& \scriptsize -
& \scriptsize 6.08430
& \scriptsize 5.6 \\
\hline
\end{tabular}
\end{center}
\end{table*}

\subsection{$\Delta^{-}$}
We here used mass $\mu=1232.0$ MeV \cite{pdg} but, in virtue of absence of 
the generally accepted experimental value of $\mu_a$ \cite{pdg} the 
value $\mu_a=$\ -3.68 from \cite{Dah} was accepted. A number of numerical 
results is given by Tables 3 and 4.
\begin{table*}
\caption{Gauge coupling constant, reduced mass $\mu_0$ and
parameters of the confining SU(3)-gluonic field for $\Delta^{-}$}
\label{t.3}
\begin{center}
\begin{tabular}{|c|c|c|c|c|c|c|c|}
\hline
\small $ g$ & \small $\mu_0$ (\small MeV) & \small $a_1$ 
& \small $a_2$ & \small $b_1$ (\small GeV) & \small $b_2$ (\small GeV) 
& \small $B_1$ & \small $B_2$ \\
\hline
\scriptsize 3.36724
& \scriptsize 2.500
& \scriptsize -0.708466
& \scriptsize 0.160912
& \scriptsize -0.187350
& \scriptsize 0.332727
& \scriptsize  -0.8100
& \scriptsize  -0.1500 \\
\hline
\scriptsize 3.36724
& \scriptsize 2.500
& \scriptsize 0.343724
& \scriptsize 0.192053
& \scriptsize -0.178990
& \scriptsize 0.362950
& \scriptsize 0.8100
& \scriptsize -0.28500 \\
\hline
\scriptsize 3.36724
& \scriptsize 2.500
& \scriptsize 0.205870
& \scriptsize -0.372948
& \scriptsize -0.254934
& \scriptsize -0.215958
& \scriptsize 0.73500
& \scriptsize -0.37500 \\
\hline
\end{tabular}
\end{center}
\end{table*}
\begin{table*}
\caption{Theoretical and experimental mass, radius and anomalous magnetic 
moment for $\Delta^{-}$}
\label{t.4}
\begin{center}
\begin{tabular}{|c|c|c|c|c|c|} 
\hline
\tiny Theoret. $\mu$ (MeV) 
&  \tiny Experim. $\mu$ (MeV) 
& \tiny Theoret. $<R>$ (fm)  
& \tiny Experim. $<R>$ (fm) 
& \tiny Theoret. $\mu_a$  
& \tiny Experim. $\mu_a$  \\
\hline
\scriptsize 1232.0
& \scriptsize $\mu= 3(m_d+\omega_j(0,0,1))= 1232.0$
& \scriptsize 0.698664
& \scriptsize -
& \scriptsize -3.68497
& \scriptsize -3.68 \\
\hline
\scriptsize 1232.0
& \scriptsize $\mu= 3(m_d+\omega_j(0,0,1))= 1232.0$
& \scriptsize 0.519021
& \scriptsize -
& \scriptsize -3.95427
& \scriptsize -3.68 \\
\hline
\scriptsize 1232.0
& \scriptsize $\mu= 3(m_d+\omega_j(0,0,1))= 1232.0$
& \scriptsize 0.358006
& \scriptsize -
& \scriptsize -3.69648
& \scriptsize -3.68 \\
\hline
\end{tabular}
\end{center}
\end{table*}

\subsection{$\Omega^-$}
In this case the generally accepted values $\mu=1672.45$ MeV 
and $\mu_a=$ -2.02 \cite{pdg} were employed. Tables 5 and 6 contain some 
numerical results.   
\begin{table*}
\caption{Gauge coupling constant, reduced mass $\mu_0$ and
parameters of the confining SU(3)-gluonic field for $\Omega^-$}
\label{t.5}
\begin{center}
\begin{tabular}{|c|c|c|c|c|c|c|c|}
\hline
\small $ g$ & \small $\mu_0$ (\small MeV) & \small $a_1$ 
& \small $a_2$ & \small $b_1$ (\small GeV) & \small $b_2$ (\small GeV) 
& \small $B_1$ & \small $B_2$ \\
\hline
\scriptsize 2.841363        
& \scriptsize 53.7500
& \scriptsize -0.753910
& \scriptsize 0.628741
& \scriptsize -0.311036
& \scriptsize -0.250568
& \scriptsize  -1.30500
& \scriptsize 1.45500 \\
\hline
\scriptsize 2.841363        
& \scriptsize 53.7500
& \scriptsize 0.231041
& \scriptsize 0.207007
& \scriptsize -0.194473
& \scriptsize 0.432236
& \scriptsize 0.6600
& \scriptsize -0.28500 \\
\hline
\scriptsize 2.841363        
& \scriptsize 53.7500
& \scriptsize -0.406736
& \scriptsize 0.369405
& \scriptsize -0.560202
& \scriptsize -0.307761
& \scriptsize -1.27500
& \scriptsize 1.15500 \\
\hline
\end{tabular}
\end{center}
\end{table*}
\newpage
\begin{table*}
\caption{Theoretical and experimental mass, radius and anomalous magnetic 
moment for $\Omega^-$}
\label{t.6}
\begin{center}
\begin{tabular}{|c|c|c|c|c|c|} 
\hline
\tiny Theoret. $\mu$ (MeV) 
&  \tiny Experim. $\mu$ (MeV) 
& \tiny Theoret. $<R>$ (fm)  
& \tiny Experim. $<R>$ (fm) 
& \tiny Theoret. $\mu_a$  
& \tiny Experim. $\mu_a$  \\
\hline
\scriptsize 1672.45 
& \scriptsize $\mu= 3(m_s+\omega_j(0,0,1))= 1672.43$
& \scriptsize 0.586809
& \scriptsize -
& \scriptsize -1.82490
& \scriptsize -2.02 \\
\hline
\scriptsize 1672.45
& \scriptsize $\mu= 3(m_s+\omega_j(0,0,1))= 1672.43$
& \scriptsize 0.430573
& \scriptsize -
& \scriptsize -2.07908
& \scriptsize -2.02 \\
\hline
\scriptsize 1672.45
& \scriptsize $\mu= 3(m_s+\omega_j(0,0,1))= 1672.43$
& \scriptsize 0.332305
& \scriptsize -
& \scriptsize -2.02136
& \scriptsize -2.02 \\
\hline
\end{tabular}
\end{center}
\end{table*}

\section{Chiral limit and concluding remarks}
\subsection{Chiral limit}
Having obtained estimates for symmetrical baryons 
we can address to the 
chiral symmetry breaking problem in QCD whose possible resolution within the 
framework of the above confinement mechanism has been discussed in 
\cite{Gon08b}. As was mentioned in subsection 2.3, merits of case consists in 
that the Dirac equation in the field (3) possesses a nontrivial spectrum of 
bound states even for massless fermions [see relation ($8^\prime$)]. As a 
result, mass of any meson remains nonzero in chiral limit when masses of quarks 
$m_q\to0$ and meson masses will be expressed only through the parameters of 
the confining SU(3)-gluonic field of (3). This purely gluonic residual mass of 
meson should be interpreted as a gluonic contribution to the meson mass.

Physically this gives us  
a possible approach to the problem of chiral symmetry breaking in QCD 
\cite{Gon08b}: in chirally symmetric world masses of mesons are fully 
determined by the confining SU(3)-gluonic field between (massless) quarks 
and not equal to zero. Accordingly chiral symmetry is a sufficiently rough 
approximation holding true only when neglecting the mentioned SU(3)-gluonic 
field between quarks and no additional mechanism of the spontaneous chiral 
symmetry breaking connected to the so-called Goldstone bosons is required. 
Referring for more details to \cite{Gon08b}, we can here only note that 
masses of symmetrical baryons have also a purely 
gluonic contribution and we may be interested in what part of their masses  
is obligatory to that contribution. 

Indeed, in chiral limit $m_{q_1}, m_{q_2}, m_{q_3}\to0$ 
we, according to (30), obtain  $(\mu)_{chiral}=3\omega$ with

$$\omega\approx\frac{g^2a_1b_1}{\lambda_1-gB_1}\approx
\frac{g^2a_2b_2}{\lambda_2-gB_2}\approx
\frac{g^2(a_1+a_2)(b_1+b_2)}{\lambda_3+g(B_1+B_2)}\ne0
\>.\eqno(35)$$

We can see that in 
chiral limit the baryon masses are completely determined only 
by the parameters $a_j, b_j, B_j$ of SU(3)-gluonic field among quarks, i.e. 
by interaction among quarks, and those masses have the purely gluonic nature. 
Accordingly, one can use the parameters $g, a_j, b_j, B_j$ adduced in Tables 1, 3, 5  
to compute $(\mu)_{chiral}$ and then find the ratio 
$gl=(\mu)_{chiral}/\mu$ which in fact represents the sought gluonic 
contribution to the baryon masses. Then, for example, for the data of 
first lines of Tables 1, 3, 5 one can easy obtain 
$gl\approx 99.254\,\%, 98.315\,\%, 72.131\,\%$, respectively for 
$\Delta^{++}$, $\Delta^{-}$, $\Omega^-$. Similar numbers hold true also 
for the other data from Tables 1, 3, 5.

So, even in chirally symmetric world, e.g., 
the symmetric baryons would have nonzero masses that would be 
determined only by SU(3)-gluonic interaction among massless quarks, i.e. 
those masses would have a purely gluonic nature. 
Moreover, since gluons are verily relativistic 
particles then the most part of masses for baryons under discussion 
is conditioned by relativistic effects, as is seen from the above estimates. 
Further discussion of the proposed chiral symmetry breaking mechanism can be 
found in \cite{{Gon08b},{Gon10}}. 

\subsection{Concluding remarks}

It should be noted that modern baryon spectroscopy in theoretical 
aspects can not be considered satisfactory enough: the main theoretical 
tools for describing dynamics of quarks in baryons here are miscellaneous 
potential models (see, e.g., early and recent reviews \cite{{BS1},{BS2}}). 
Quark interaction at such a description is modelled mainly by either 
oscillator potential or the 
confining potential of form $a/r+br$. But, as follows from the uniqueness 
theorem of Section 2 (see also Appendix D), such potentials cannot describe any gluon 
configuration between quarks. It would be possible if the mentioned potentials were  
solutions of Yang-Mills equations directly derived from QCD-Lagrangian since, 
from the QCD-point of view, any gluonic field should be a solution of 
Yang-Mills equations (as well as any electromagnetic field is by definition 
always a solution of Maxwell equations). According to the above uniqueness 
theorem neither oscillator potential nor the 
confining potential of form $a/r+br$ are not solutions of Yang-Mills 
equations. In particular, Coulomb and linear parts belong to different parts of the 
YM-potentials (see (3)), accordingly, to the colour electric and colour magnetic parts 
so they cannot be united in one component of form $a/r+br$.
So, we draw the conclusion 
(mentioned as far back as in Refs. \cite{{Gon031},{Gon032}} and elaborated 
more in detail in Ref. \cite{Gon08a}) that the potential approach seems to be 
inconsistent. 

Another matter is our confinement mechanism proposed in 
\cite{{Gon01},{Gon051},{Gon052}} 
which gives new possibilities from the first 
principles of QCD immediately appealing 
to the quark and gluonic degrees of freedom. Indeed, the words 
{\em quark and gluonic degrees of freedom} make exact sense: gluons come 
forward in the form of bosonic condensate described by parameters $a_j$, 
$b_j$, $B_j$ from the unique exact solution (3) of the Yang-Mills equations 
while quarks are represented by their current masses $m_q$.

Consequently, the given mechanism is based on the unique family of compatible 
nonperturbative solutions for the Dirac-Yang-Mills system directly derived 
from QCD-Lagrangian and, as a result, the approach is itself nonperturbative, 
relativistic from the outset, admits self-consistent nonrelativistic limit 
and may be employed for any meson (quarkonium). Applications of the approach  
in Refs. 
\cite{{Gon031},{Gon032},{Gon04},{Gon08a},{Gon06},{Gon07a},{Gon07b},{Gon08},{Gon08b},{Gon10}}
allow one to speak about the fact that the 
confinement mechanism elaborated in \cite{{Gon01},{Gon051},{Gon052}} 
gives new possibilities for considering many old problems of hadronic 
(meson) physics (such as nonperturbative computation of decay constants, masses 
and radii of mesons, chiral symmetry breaking, the origin of the so-called 
scale $\Lambda_{QCD}$ \cite{Gon12a} and so forth) from the first 
principles of QCD immediately appealing to the quark and gluonic degrees of freedom.

At the same time, as the 
results of the present Chapter show, this mechanism may be extended over baryons 
in a reasonable way. We hope further study of baryons with the help of 
the given approach to shed new light on the internal structure of baryons. 

\section{Problem of masses in particle physics}
\subsection{Preliminaries}
As is known \cite{pdg}, the generally accepted standard model with one 
Higgs doublet asserts that the masses of fundamental fermions (quarks and 
leptons) are acquired through the Higgs mechanism so for their   
masses $m_i$ we obtain (without taking mixings into account) 
$m_i=f_{i}v/\sqrt{2}$, where the vacuum Higgs condensate $v\approx246$ GeV and 
$i$ stands for quark and lepton flavours. But little is known about the coupling 
constants $f_{i}$ and much may be elucidated 
only with discovering Higgs bosons. The same holds true for the gauge bosons 
$W^{\pm}$, $Z$ where masses $m_W=ev/(2\sin{\theta_W})$, $m_Z=ev/(\sin{2\theta_W})$ with 
the so-called weak angle $\theta_W$ so that 
$\sin^2{\theta_W}\approx0.23$ and $e$ is the elementary electric charge. If 
taking into account that the mass of Higgs boson $m_H=\lambda v$ with 
a self-interaction constant $\lambda$ then it is clear that masses of all the 
abovementioned particles are proportional to $m_H$, and, consequently, the 
discovery of Higgs boson will not completely resolve the puzzle of origin 
of masses in particle physics -- the question will remain where the mass 
$m_H$ comes from not speaking already about the nature of the above 
miscellaneous constants $f_i$ and $\lambda$. 

At present, to our mind, one can single out two most promising approaches to 
a possible resolution for the mentioned problems: technicolour theories and 
preon models. Under the circumstances let us shortly outline how both these 
directions might be estimated from the point of view of our confinement 
mechanism and the chiral symmetry breaking one based on the latter  
and discussed above and in \cite{Gon08b}. 
\subsection{Technicolour theories}
Referring for more details concerning those models to both early references 
\cite{{T11},{T12},{T13}} and modern status of them (see, e.g., \cite{T2}) let 
us note the following. The main idea of acquiring masses, e.g., for $W^{\pm}$ and $Z$ 
bosons, consists in that a new set of the so-called techniquarks is postulated 
at the energy scale of order 1 TeV which interact with each other through 
the technigluons and it makes the massless techipions exist as Goldstone bosons. 
The latter give masses to $W^{\pm}$ and $Z$ after spontaneous symmetry breaking. 
It should be noted, however, those massless technipions appear as a result of 
violating chiral symmetry connected with technicolour QCD on the analogy with 
chiral symmetry breaking in usual QCD. But, as we have discussed in 
\cite{Gon08b} and in Section 6, the hypothetical mechanism for chiral symmetry 
breaking in standard QCD with appearance of Goldstone bosons (pions) seems to 
fail because of pions can never be massless inasmuch as they have nonzero masses 
even in chirally symmetric world due to gluons. The same will also perfectly 
hold true for technipions which would always have nonzero masses due to 
technigluons since technicolour QCD should manifest the confinement mechanism 
similar to our one in usual QCD. Therefore, technicolour theories look rather 
doubtful from the point of view of our confinement mechanism. 
\subsection{Preon models}
Another cardinal approach to the problem of masses is connected with the 
preon models (see, e.g., \cite{Pr} and references therein). Under this approch 
quarks, leptons and gauge vector bosons are suggested to be composed of stable 
spin-1/2 preons, for example, existing in three flavours and being combined 
according to simple rules. The main theoretical objection to preon theories is 
the mass paradox which arises by virtue of the Heisenberg's uncertainty 
principle. Scattering experiments have shown \cite{Per} that quarks and 
leptons are point-like up to the scales of order $10^{-3}$ fm which corresponds to 
a preon mass of order 197 GeV (due to the uncertainty 
principle) if the preon is confined to a box of 
such a size, i.e. its mass will approximately be $0.4\times10^{5}$ times greater 
than, e.g., that of $d$-quark. 
Thus, the preon models are faced with a mass paradox: how could quarks or electrons 
be made of smaller particles that would have masses of many orders of magnitude 
greater than the fundamental fermion masses? The paradox might be resolved by 
the rather dubious postulate about a large binding force between preons 
cancelling their mass-energies. Our confinement mechanism points out the more 
physically acceptable way of overcoming these obstacles. If the interaction
among preons is decribed by a QCD-like theory based on, e.g., SU(N)-group with 
$N\ge2$ then, according to our results \cite{Gon051,Gon052}, such theories should 
also manifest confinement to generate masses decribed by relations similar 
to (8) and $(8^\prime)$. This signifies that preons might possess small masses 
or be just massless and, as a result, mass paradox would be removed. 

Among fundamental fermions one should pay special attention to electron. Within the 
framework of preon models the most natural composition for it should be the one of three 
preons \cite{Pr}, so electron might be described by triplet $(\beta,\beta,\delta)$, 
where preons $\beta$ and $\delta$ carry electric charges $-2e/3$, $e/3$,  
respectively, with elementary charge $e$ \cite{Pr}. If supposing that interaction among 
them is desribed by the gauge 
group SU(3) through pre-gluons \cite{Pr} then we formally come to the same scheme which 
was under exploration for baryons in the given chapter. Under the circumstances we can 
find the parameters in solution (3) which will correspond to the observed mass, some 
radius (of order of the classical electron radius $r_e\sim10^{15}$ m) and (anomalous) 
magnetic moment for electron. In those calculations one can consider the mentioned 
preons to be massless, as is said above. We hope to study this interesting task 
elsewhere. 
\section{Remarks about the Yang-Mills Millennium problem}

Our approach to the quark confinement mechanism is closely connected with the so-called 
Yang-Mills Millennium problem (see \cite{MP} for more details). Its formulation looks as 
follows:
\vskip0.5cm
\vbox{{\bf Yang-Mills Existence and Mass Gap}: {\em Prove that for any
compact simple gauge group $G$, quantum Yang-Mills theory on ${\mathbb R}^4$
exists and has a mass gap $\Delta > 0$.}}
\vskip0.5cm
It is not complicated to see that our results on the confinement mechanism [in 
particular, the solution (3)] to a certain degree clear the way to obtain a proof of the 
above problem. So long as group SU(3) belongs to the class of groups $G$ from the 
formulation of problem then let us demonstrate possible opportunities with an example of 
the SU(3) gauge group and then make some remarks concerning the common case. 

Let us note the following. From mathematical 
point of view the Yang-Mills field are the {\em connections} in vector (or, which is 
equivalent, in principal) bundles over some manifold $M$. When $M$ is the Minkowski space 
the situation is simplified so long as all the bundles over $M$ with topology 
${\mathbb R}^4$ are trivial. When quantising Yang-Mills field we obtain, from physical 
point of view, quanta of that field, e.g., photons or gluons, which realize an 
interaction among the corresponding (matter) particles able to interchange by those 
quanta. But from mathematical point of view the mentioned particles are described by the 
{\em cross-sections} of vector bundles over the given spacetime, in particular, over 
${\mathbb R}^4$. As a result, mathematically, connections do not live separately from 
cross-sections while, physically, the sources of any quanta are the matter particles. 

Under the circumstances when discussing the above millennium problem we should take into 
account also the matter fields rather than restrict ourselves to the pure Yang-Mills 
theory and further we should pass on to the pure Yang-Mills case by some limiting 
procedure. Then, returning to group SU(3), we have seen that there exists the exact 
solution (3) of the Yang-Mills equations which is unique in a certain sense (see 
Appendix D) and spectrum of bound states in such a field is given by the relation (8). 
The latter formula cannot be obtained without including spinor fields as cross-sections 
of the corresponding bundle into consideration. 

We now especially are interested in the limiting case of (8), namely in the relation 
$(8^\prime)$. In it all the information about matter (spinor) fields vanished since that 
information was related only with mass parameters in the conforming Dirac equation. As a 
consequence, the relation $(8^\prime)$ should be interpreted as the pure gluonic energy.  
Obviously, the latter only depends on quantities $a_j$, $b_j$, $B_j$ created by 
group SU(3) and is quantized in virtue of availability of the nonnegative integer 
numbers $n_j$. Also 
from considerations of Section 2 it is clear that the next relation should be fulfilled 
$$(\omega_1)_{\rm chiral}=(\omega_2)_{\rm chiral}=(\omega_3)_{\rm chiral},\eqno(36)$$
which gives an additional number of restrictions on the parameters $a_j$, $b_j$, $B_j$. 
Further we can see that there is a mass gap $\Delta$ defined at $n_j=0$, $l_j=0$ 
from the relation
$$\Delta=\frac{a_1b_1}{\pm1-gB_1}=\frac{a_2b_2}{\pm1-gB_2}=
\frac{(a_1+a_2)(b_1+b_2)}{\pm1+g(B_1+B_2)}.\eqno(37)$$

Of course, $\Delta$ can be positive or negative depending from a choice of parameters 
$a_j$, $b_j$, $B_j$ but parameters $a_j$, $b_j$ should not be equal to 0. I. e., we 
obtained a quantum theory for SU(3)-group asserting that any reasonable energy connected 
with its quanta (gluons) is quantized and has a mass gap. As an application of those  
properties we have already discussed the chiral symmetry breaking mechanism in 
Sections 2 and 6 and in Refs. \cite{{Gon08b},{Gon10}}. 

Passing on to the general case, we can note that (see Appendix D) 
the obtained results may be extended 
over all semisimple compact Lie groups since for them the corresponding Lie 
algebras possess just the only Cartan subalgebra. Also we can talk about the 
compact non-semisimple groups, for example, U($N$). In the latter case 
additionally to Cartan subalgebra we have centrum consisting from the matrices 
of the form $\alpha I_N$ ($I_N$ is the unit matrix $N\times N$) with arbitrary 
constant $\alpha$. For all those groups there exists their own uniqueness theorem, 
as follows from considerations in Appendix D, one can evaluate spectrum of bound states 
(examples for groups SU(2) and SU(4) can be found in \cite{Gon052}) and chiral limit that 
entails results about quantizing energy of the corresponding quanta and gives a mass gap 
for all those groups. To get the formulas in an explicit form one should take a 
concrete realization of the conforming Lie algebras. 

The most relevant physical cases are of course U(1)- and SU(3)-ones 
(QED and QCD). In particular, the U(1)-case allows us to build the classical 
model of confinement (see Section 2 and Ref. \cite{GF10}).

\section{Conclusion}

The main idea of quark confinement may be borrowed from classical electrodynamics. 
Indeed,  let us recall the well-known case of 
motion of a charged particle in the homogeneous magnetic field 
(see, e.g., Ref. \cite{LL}). In the latter case the particle moves along helical curve 
with lead of helix $h=2\pi mv\cos{\alpha}/(qH\sqrt{1-v^2})$ and radius 
$R=mv\sin{\alpha}/(qB\sqrt{1-v^2})$, where $\alpha$ is an angle between vectors of 
the particle velocity $\bf v$ and magnetic induction $\bf B$, $q$ is particle charge, 
$m$ is particle mass. As a consequence, the homogeneous magnetic 
field does not give rise to the full confinement of the particle since the 
latter may go to infinity along the helical curve. The situation is not changed 
at quantum level as well: there exist no bound states in the homogeneous 
magnetic field \cite{LLQ}. But if estimating module of $B$ at 
$R\sim10^{-15}\,{\rm m}=1$ fm for electron then $B$ will be of order $10^{23}$ T. I. e., 
if considering that quarks are confined by a colour magnetic field that they themselves 
create then one needs a colour magnetic field between quarks with $B$ of such an order  
and that field should not allow to quark to go to infinity. It is clear this field should 
be a solution of the Yang-Mills equations. Really, if speaking about Minkowski spacetime 
then searching for classical solutions of the Yang-Mills equations makes sense because 
at large distances between quarks the latter are surrounded with a huge number of gluons 
that are emitted by both quarks and gluons themselves. Under this situation it is quite 
plausible that confinement of quarks arises due to certain properties of such gluonic 
clouds while the latter should be described just by {\em classical} solutions of 
the Yang-Mills equations. 

As was shown in Refs. \cite{{Gon01},{Gon051},{Gon052}}, the necessary solution has the 
form (3) for group SU(3) and is unique in a certain sense. 
This result allowed us to propose a quark confinement mechanism which was successfully 
applied to meson spectroscopy. Indeed, as follows from (16) at $|b_j|\to\infty$ we 
have $<r>\,\sim\, \sqrt{\sum\limits_{j=1}^3\frac{1}
{(g|b_j|)^2}}$, so in the strong magnetic colour field when $|b_j|\to\infty$, 
$<r>\to 0$, while the meson wave functions of (6) and (9) behave as 
$\Psi_j\,\sim\,e^{-g|b_j|r}$, i. e., just the magnetic colour field of (3) 
provides two quarks with confinement. This situation also holds true at classical 
level \cite{GF10}.

The given Chapter extends this mechanism over baryons and further 
study of that approach will be connected with analysis of concrete baryons with 
its help.

\section{Appendix A}
We here represent some results about eigenspinors of the Euclidean Dirac 
operator on two-sphere ${\mathbb S}^2$ employed in the main part of the Chapter. 

When separating variables in the Dirac equation (4) there naturally 
arises the Euclidean Dirac operator ${\cal D}_0$ on the unit two-dimensional 
sphere ${\mathbb S}^2$ and we should know its eigenvalues with the corresponding 
eigenspinors. Such a problem also arises in the black hole theory while 
describing the so-called twisted spinors on Schwarzschild and 
Reissner-Nordstr\"om black holes and it was analysed in 
Refs. \cite{{Gon052},{Gon99a},{Gon99b}}, so we can use the results obtained 
therein for our aims. Let us adduce the necessary relations. 

The eigenvalue equation for
corresponding spinors $\Phi$ may look as follows
$${\cal D}_0\Phi=\lambda\Phi.\>\eqno({\rm A.1})$$

As was discussed in Refs. \cite{{Gon99a},{Gon99b}} 
the natural form of ${\cal D}_0$ 
(arising within applications) in 
local coordinates $\vartheta, \varphi$ on the unit sphere ${\mathbb S}^2$ looks 
as 
$${\cal D}_0=-i\sigma_1\left[
i\sigma_2\partial_\vartheta+i\sigma_3\frac{1}{\sin{\vartheta}}
\left(\partial_\varphi-\frac{1}{2}\sigma_2\sigma_3\cos{\vartheta}
\right)\right]=$$
$$\sigma_1\sigma_2\partial_\vartheta+\frac{1}{\sin\vartheta}
\sigma_1\sigma_3\partial_\varphi+ \frac{\cot\vartheta}{2}
\sigma_1\sigma_2         \eqno(\rm A.2)$$
with the ordinary Pauli matrices
$$\sigma_1=\pmatrix{0&1\cr 1&0\cr}\,,\sigma_2=\pmatrix{0&-i\cr i&0\cr}\,,
\sigma_3=\pmatrix{1&0\cr 0&-1\cr}\,, $$
so that $\sigma_1{\cal D}_0=-{\cal D}_0\sigma_1$.

The equation (A.1) was explored in Refs. \cite{{Gon99a},{Gon99b}}.
Spectrum of $D_0$ consists of the numbers
$\lambda=\pm(l+1)$              
with multiplicity $2(l+1)$ of each one, where $l=0,1,2,...$. Let us 
introduce the number $m$ such that $-l\le m\le l+1$ and the corresponding 
number $m'=m-1/2$ so $|m'|\le l+1/2$. Then the conforming eigenspinors of  
operator ${\cal D}_0$ are 
$$\Phi=\pmatrix{\Phi_1\cr\Phi_2\cr}= 
\Phi_{\mp\lambda}=\frac{C}{2}\pmatrix{P^k_{m'-1/2}\pm P^k_{m'1/2}\cr
P^k_{m'-1/2}\mp P^k_{m'1/2}\cr}e^{-im'\varphi}\> \eqno(\rm A.3) $$
with the coefficient $C=\sqrt{\frac{l+1}{2\pi}}$ and $k=l+1/2$.  
These spinors form an orthonormal basis in $L_2^2({\mathbb S}^2)$ 
and are subject 
to the normalization condition
$$\int_{{\mathbb S}^2}\Phi^{\dag}\Phi d\Omega=
\int\limits_0^\pi\,\int\limits_0^{2\pi}(|\Phi_{1}|^2+|\Phi_{2}|^2)
\sin\vartheta d\vartheta d\varphi=1\>. \eqno(\rm A.4)$$
Further, owing to the relation $\sigma_1{\cal D}_0=-{\cal D}_0\sigma_1$ we, 
obviously, have
$$ \sigma_1\Phi_{\mp\lambda}=\Phi_{\pm\lambda}\,.  \eqno(\rm A.5)$$

As to functions $P^k_{m'n'}(\cos\vartheta)\equiv P^k_{m',\,n'}(\cos\vartheta)$ 
then they can be chosen by 
miscellaneous ways, for instance, as follows (see, e. g.,
Ref. \cite{Vil91})
$$P^k_{m'n'}(\cos\vartheta)=$$
$$i^{-m'-n'}
\sqrt{\frac{(k-m')!(k-n')!}{(k+m')!(k+n')!}}
\left(\frac{1+\cos{\vartheta}}{1-\cos{\vartheta}}\right)^{\frac{m'+n'}{2}}\,
\times$$
$$\times\sum\limits_{j={\rm{max}}(m',n')}^k
\frac{(k+j)!i^{2j}}{(k-j)!(j-m')!(j-n')!}
\left(\frac{1-\cos{\vartheta}}{2}\right)^j \eqno(\rm A.6)$$
with the orthogonality relation at $m',n'$ fixed
$$\int\limits_0^\pi\,{P^{*k}_{m'n'}}(\cos\vartheta)
P^{k'}_{m'n'}(\cos\vartheta)
\sin\vartheta d\vartheta={2\over2k+1}\delta_{kk'}
\>.\eqno(\rm A.7)$$
It should be noted that square of 
${\cal D}_0$ is 
$${\cal D}^2_0=-\Delta_{{\mathbb S}^2}I_2+
\sigma_2\sigma_3\frac{\cos{\vartheta}}{\sin^2{\vartheta}}\partial_\varphi
+\frac{1}{4\sin^2{\vartheta}} +\frac{1}{4}\>,
\eqno(\rm A.8)$$
while laplacian on the unit sphere is
$$\Delta_{{\mathbb S}^2}=
\frac{1}{\sin{\vartheta}}\partial_\vartheta\sin{\vartheta}\partial_\vartheta+
\frac{1}{\sin^2{\vartheta}}\partial^2_\varphi=
\partial^2_\vartheta+\cot{\vartheta}\partial_\vartheta
+\frac{1}{\sin^2{\vartheta}}\partial^2_\varphi\>,
\eqno(\rm A.9)$$
so the relation (A.8) is a particular case of the so-called 
Weitzenb{\"o}ck-Lichnerowicz formulas (see Refs. \cite{81}). 
Then from (A.1) it follows 
${\cal D}^2_0\Phi=\lambda^2\Phi$ and, when using the ansatz  
$\Phi=P(\vartheta)e^{-im'\varphi}=\pmatrix{P_1\cr P_2\cr}e^{-im'\varphi}$, 
$P_{1,2}=P_{1,2}(\vartheta)$, the equation ${\cal D}^2_0\Phi=\lambda^2\Phi$ 
turns into 
$$\left(-\partial^2_\vartheta-\cot{\vartheta}\partial_\vartheta +
\frac{m'^2+\frac{1}{4}}{\sin^2{\vartheta}}+
\frac{m'\cos{\vartheta}}{\sin^2{\vartheta}}\sigma_1\right)P=$$
$$\left(\lambda^2-\frac{1}{4}\right)P\>,
\eqno(\rm A.10)$$
wherefrom all the above results concerning spectrum of ${\cal D}_0$ can be 
derived \cite{{Gon99a},{Gon99b}}.

When calculating the functions $P^k_{m'n'}(\cos\vartheta)$ directly, to our 
mind, it is the most convenient to use the integral expression \cite{Vil91}

$$P^k_{m'n'}(\cos\vartheta)=\frac{1}{2\pi}
\sqrt{\frac{(k-m')!(k+m')!}{(k-n')!(k+n')!}}$$
$$\int_{0}^{2\pi}\left(e^{i\varphi/2}\cos{\frac{\vartheta}{2}}+
ie^{-i\varphi/2}\sin{\frac{\vartheta}{2}}\right)^{k-n'}\times$$
$$\left(ie^{i\varphi/2}\sin{\frac{\vartheta}{2}}+
e^{-i\varphi/2}\cos{\frac{\vartheta}{2}}\right)^{k+n'}e^{im'\varphi}d\varphi 
\eqno(\rm A.11)$$
and the symmetry relations ($z=\cos{\vartheta}$) 
$$P^k_{m'n'}(z)=P^k_{n'm'}(z), \>P^k_{m',-n'}(z)=P^k_{-m',\,n'}(z),$$ 
$$P^k_{m'n'}(z)=P^k_{-m',-n'}(z)\,,$$ 
$$P^k_{m'n'}(-z)=i^{2k-2m'-2n'}P^k_{m',-n'}(z)\>. \eqno(\rm A.12)$$
In particular
$$P^{k}_{kk}(z)=
\cos^{2k}{(\vartheta/2)},  
P^{k}_{k,-k}(z)=i^{2k}\sin^{2k}{(\vartheta/2)},$$
$$P^{k}_{k0}(z)=\frac{i^{k}\sqrt{(2k)!}}{2^k k!}\sin^{k}{\vartheta}\,,$$
$$ P^{k}_{kn'}(z)=i^{k-n'}\sqrt{\frac{(2k)!}{(k-n')!(k+n')!}}\times$$
$$\sin^{k-n'}{(\vartheta/2)}\cos^{k+n'}{(\vartheta/2)}\>. \eqno(\rm A.13)$$ 
\subsection*{Eigenspinors with $\lambda=\pm1,\,\pm2$}
If $\lambda=\pm(l+1)=\pm1$ then $l=0$ and from (A.3) it follows that 
$k=l+1/2=1/2$, $|m'|\le1/2$ and we need the functions $P^{1/2}_{m',\pm1/2}$ 
that are easily evaluated with the help of (A.11)--(A.13) so   
the eigenspinors for $\lambda=-1$ are 
$$\Phi=\frac{C}{2}\pmatrix{\cos{\frac{\vartheta}{2}}+
i\sin{\frac{\vartheta}{2}}\cr
\cos{\frac{\vartheta}{2}}-i\sin{\frac{\vartheta}{2}}\cr}e^{i\varphi/2},$$
$$\Phi=\frac{C}{2}\pmatrix{\cos{\frac{\vartheta}{2}}+
i\sin{\frac{\vartheta}{2}}\cr
-\cos{\frac{\vartheta}{2}}+i\sin{\frac{\vartheta}{2}}\cr}
e^{-i\varphi/2},\eqno(\rm A.14)$$
while for $\lambda=1$ the conforming spinors are
$$\Phi=\frac{C}{2}\pmatrix{\cos{\frac{\vartheta}{2}}-
i\sin{\frac{\vartheta}{2}}\cr
\cos{\frac{\vartheta}{2}}+i\sin{\frac{\vartheta}{2}}\cr}e^{i\varphi/2},$$ 
$$\Phi=\frac{C}{2}\pmatrix{-\cos{\frac{\vartheta}{2}}+
i\sin{\frac{\vartheta}{2}}\cr
\cos{\frac{\vartheta}{2}}+i\sin{\frac{\vartheta}{2}}\cr}e^{-i\varphi/2}
\eqno(\rm A.15) $$
with the coefficient $C=\sqrt{1/(2\pi)}$.

It is clear that (A.14)--(A.15) can be rewritten in the form 
$$\lambda=-1: \Phi=\frac{C}{2}\pmatrix{e^{i\frac{\vartheta}{2}}
\cr e^{-i\frac{\vartheta}{2}}\cr}e^{i\varphi/2},$$
or
$$\Phi=\frac{C}{2}\pmatrix{e^{i\frac{\vartheta}{2}}\cr
-e^{-i\frac{\vartheta}{2}}\cr}e^{-i\varphi/2},$$
$$\lambda=1: \Phi=\frac{C}{2}\pmatrix{e^{-i\frac{\vartheta}{2}}\cr
e^{i\frac{\vartheta}{2}}\cr}e^{i\varphi/2},$$
or
$$\Phi=\frac{C}{2}\pmatrix{-e^{-i\frac{\vartheta}{2}}\cr
e^{i\frac{\vartheta}{2}}\cr}e^{-i\varphi/2}\,, 
\eqno(\rm A.16) $$
so the relation (A.5) is easily verified at $\lambda=\pm1$. 

In studying vector mesons and excited states of heavy quarkonia eigenspinors 
with $\lambda=\pm2$ may also be useful. Then $k=l+1/2=3/2$, $|m'|\le3/2$ and we 
need the functions $P^{3/2}_{m',\pm1/2}$ 
that can be evaluated with the help of (A.11)--(A.13). Computation gives 
rise to 
$$ P^{3/2}_{3/2,-1/2}=-\frac{\sqrt{3}}{2}\sin{\vartheta}
\sin{\frac{\vartheta}{2}}= P^{3/2}_{-3/2,1/2},\>$$   
$$P^{3/2}_{3/2,1/2}=i\frac{\sqrt{3}}{2}\sin{\vartheta}
\cos{\frac{\vartheta}{2}}= P^{3/2}_{-3/2,-1/2},\>$$
$$P^{3/2}_{1/2,-1/2}= -\frac{i}{4}\left(\sin{\frac{\vartheta}{2}}-
3\sin{\frac{3}{2}\vartheta}\right)=P^{3/2}_{-1/2,1/2},\>$$
$$P^{3/2}_{1/2,1/2}= \frac{1}{4}\left(\cos{\frac{\vartheta}{2}}+
3\cos{\frac{3}{2}\vartheta}\right)=P^{3/2}_{-1/2,-1/2},\>
\eqno(\rm A.17) $$
and according to (A.3) this entails eigenspinors with $\lambda=2$ in the 
form
$$\frac{C}{2}i\frac{\sqrt{3}}{2}\sin{\vartheta}
\pmatrix{e^{-i\frac{\vartheta}{2}}\cr
e^{i\frac{\vartheta}{2}}\cr}e^{i3\varphi/2},\>
\frac{C}{8}\pmatrix{3e^{-i\frac{3\vartheta}{2}}+e^{i\frac{\vartheta}{2}}\cr
3e^{i\frac{3\vartheta}{2}}+e^{-i\frac{\vartheta}{2}}\cr}e^{i\varphi/2},\>$$
$$\frac{C}{8}\pmatrix{-3e^{-i\frac{3\vartheta}{2}}-e^{i\frac{\vartheta}{2}}\cr
3e^{i\frac{3\vartheta}{2}}+e^{-i\frac{\vartheta}{2}}\cr}e^{-i\varphi/2},\>
\frac{C}{2}i\frac{\sqrt{3}}{2}\sin{\vartheta}
\pmatrix{-e^{-i\frac{\vartheta}{2}}\cr
e^{i\frac{\vartheta}{2}}\cr}e^{-i3\varphi/2}\>
 \eqno(\rm A.18) $$
with $C=1/\sqrt{\pi}$, while eigenspinors with $\lambda=-2$ are obtained in 
accordance with relation (A.5). 

\section{Appendix B}
We here adduce the explicit form for the radial parts of meson wave functions 
from (6). At $n_j=0$ they are given by 
$$F_{j1}=C_jP_jr^{\alpha_j}e^{-\beta_jr}\left(1-
\frac{Y_j}{Z_j}\right),$$
$$F_{j2}=iC_jQ_jr^{\alpha_j}e^{-\beta_jr}\left(1+
\frac{Y_j}{Z_j}\right),\eqno(\rm B.1)$$
while at $n_j>0$ they are given by
$$F_{j1}=C_jP_jr^{\alpha_j}e^{-\beta_jr}\times$$
$$\left[\left(1-\frac{Y_j}{Z_j}\right)L^{2\alpha_j}_{n_j}(r_j)+
\frac{P_jQ_j}{Z_j}r_jL^{2\alpha_j+1}_{n_j-1}(r_j)\right],$$
$$F_{j2}=iC_jQ_jr^{\alpha_j}e^{-\beta_jr}\times$$
$$\left[\left(1+
\frac{Y_j}{Z_j}\right)L^{2\alpha_j}_{n_j}(r_j)-
\frac{P_jQ_j}{Z_j}r_jL^{2\alpha_j+1}_{n_j-1}(r_j)\right]\eqno(\rm B.2)$$
with the Laguerre polynomials $L^\rho_{n}(r_j)$, $r_j=2\beta_jr$, 
$\beta_j=\sqrt{\mu_0^2-\omega_j^2+g^2b_j^2}$ at $j=1,2,3$ with 
$b_3=-(b_1+b_2)$, 
$P_j=gb_j+\beta_j$, $Q_j=\mu_0-\omega_j$,
$Y_j=P_jQ_j\alpha_j+(P^2_j-Q^2_j)ga_j/2$, 
$Z_j=P_jQ_j\Lambda_j+(P^2_j+Q^2_j)ga_j/2$    
with $a_3=-(a_1+a_2)$,   
$\Lambda_j=\lambda_j-gB_j$ with $B_3=-(B_1+B_2)$, 
$\alpha_j=\sqrt{\Lambda_j^2-g^2a_j^2}$, 
while $\lambda_j=\pm(l_j+1)$ are
the eigenvalues of Euclidean Dirac operator ${\cal D}_0$ 
on unit two-sphere with $l_j=0,1,2,...$ (see Appendix A) 
and quantum numbers $n_j=0,1,2,...$ are defined by the relations 
$$n_j=\frac{gb_jZ_j-\beta_jY_j}{\beta_jP_jQ_j}\,, 
\eqno(\rm B.3)$$
which entails the spectrum (8).  
Further, $C_j$ of (B.1)--(B.2) should be determined
from the normalization condition
$$\int_0^\infty(|F_{j1}|^2+|F_{j2}|^2)dr=\frac{1}{3}\>.\eqno(\rm B.4)$$
As a consequence, we shall gain that in (4) 
$\Psi_j\in L_2^{4}({\mathbb R}^3)$ at any $t\in{\mathbb R}$ and, accordingly,
$\Psi=(\Psi_1,\Psi_2,\Psi_3)$ may describe relativistic bound states 
in the field (3) with the energy spectrum (8). As is clear from (B.3) at 
$n_j=0$ we have 
$gb_j/\beta_j=Y_j/Z_j$ so the radial parts of (B.1) can be rewritten as  
$$F_{j1}=C_jP_jr^{\alpha_j}e^{-\beta_jr}\left(1-
\frac{gb_j}{\beta_j}\right),$$
$$F_{j2}=iC_jQ_jr^{\alpha_j}e^{-\beta_jr}\left(1+
\frac{gb_j}{\beta_j}\right)\>.\eqno(\rm B.5)$$
More details can be found in Refs. \cite{{Gon01},{Gon052}}. 

\section{Appendix C}
We here recall some facts about tensorial products of Hilbert spaces
necessary in the main part of the Chapter. For more information see, e.g., 
Ref. \cite{Vil91} and references therein. 

For two Hilbert spaces ${\mathcal H}_1$ and ${\mathcal H}_2$, by definition, 
their tensorial product ${\mathcal H}_1\otimes{\mathcal H}_2$ 
consists from every possible linear combinations of elements of the form 
$\vec x_1\otimes\vec x_2$, where $\vec x_1\in{\mathcal H}_1$, 
$\vec x_2\in{\mathcal H}_2$. The obvious relations take place 
$$(\alpha\vec x_1+\beta\vec x_2)\otimes\vec y=\alpha(\vec x_1\otimes\vec y)+
\beta(\vec x_2\otimes\vec y)\>,          \eqno(\rm C.1)$$
$$x\otimes(\alpha\vec y_1+\beta\vec y_2)=\alpha(\vec x\otimes\vec y_1)+
\beta(\vec x\otimes\vec y_2)          \eqno(\rm C.2)$$
with arbitrary complex numbers $\alpha$, $\beta$. 

Scalar product in ${\mathcal H}_1\otimes{\mathcal H}_2$ is defined by the 
equality
$$(\vec x_1\otimes\vec y_1,\vec x_2\otimes\vec y_2)=
(x_1,x_2)_1(y_1,y_2)_2\>, \eqno(\rm C.3) $$
where $(x_1,x_2)_1$, $(y_1,y_2)_2$ are the scalar products, respectively, in 
${\mathcal H}_1$ and ${\mathcal H}_2$. Under the situation, if $\{\vec e_i\}$ 
is an orthonormal basis in ${\mathcal H}_1$ and $\{\vec f_j\}$ 
is an orthonormal basis in ${\mathcal H}_2$ then 
$\vec h_{ij}=\vec e_i\otimes\vec f_j$ is an orthonormal basis in 
${\mathcal H}_1\otimes{\mathcal H}_2$. 

Further, if $A$ is a linear operator in ${\mathcal H}_1$ and $B$ is a linear 
operator in ${\mathcal H}_2$ then tensorial (Kronecker) product $A\otimes B$ 
acts in 
${\mathcal H}_1\otimes{\mathcal H}_2$ and is defined by 
relation $(A\otimes B)(\vec x_1\otimes \vec x_2)=A\vec x_1\otimes B\vec x_2$ 
with the properties
$$A\otimes(\alpha B_1+\beta B_2)=\alpha A\otimes B_1+
\beta A\otimes B_2\>,                        \eqno(\rm C.4)$$
$$(\alpha A_1+\beta A_2)\otimes B=\alpha A_1\otimes B+
\beta A_2\otimes B\>,                        \eqno(\rm C.5)$$
$$A_1A_2\otimes B_1B_2=(A_1\otimes B_1)
(A_2\otimes B_2)\>.                        \eqno(\rm C.6)$$

In a similar way, the Kronecker sum of $A$ and $B$ is defined by 
$A\oplus B=A\otimes I +I\otimes B$ with identical operator $I$. Under the 
circumstances, let $\lambda_i$ be the eigenvalues $A$ and $\mu_j$ be those 
of $B$ (listed according to multiplicity). Then the eigenvalues of 
$A\otimes B$ are $\lambda_i\mu_j$ while those of $A\oplus B$ will be 
$\lambda_i+\mu_j$. 

All the above is directly generalized to the arbitrary number of Hilbert 
spaces. For example, the operator 
$A_1\otimes I\otimes I+I\otimes A_2\otimes I+I\otimes I\otimes A_3$ has the 
eigenvalues $\lambda_i+\mu_j+\nu_k$ in space 
${\mathcal H}_1\otimes{\mathcal H}_2\otimes{\mathcal H}_3$ if $\lambda_i$, 
$\mu_j$, $\nu_k$ are the eigenvalues of $A_1$, $A_2$ and $A_3$, respectively, 
in ${\mathcal H}_1$, ${\mathcal H}_2$, ${\mathcal H}_3$. 

\section{Appendix D}
The facts adduced here have been obained in Refs. \cite{{Gon051},{Gon052}} and 
we concisely give them only for completeness of discussion in Section 2.

To specify the question, let us note that in general the Yang-Mills equations 
on a manifold $M$ can be written as
$$d\ast F= g(\ast F\wedge A - A\wedge\ast F) \>,\eqno(\mathrm D.1)$$ 
where a gluonic field $A=A_\mu dx^\mu=
A^a_\mu \lambda_adx^\mu$ [$\lambda_a$ are the 
known Gell-Mann matrices, $\mu=t,r,\vartheta,\varphi$ (in the case of 
spherical coordinates), $a=1,...,8$], 
the curvature matrix (field strentgh)
$F=dA+gA\wedge A= F^a_{\mu\nu}\lambda_adx^\mu\wedge dx^\nu$ with exterior 
differential $d$ and the Cartan's (exterior) product $\wedge$, while $\ast$ 
means the Hodge star operator conforming to a metric on manifold under 
consideration, $g$ is a gauge coupling constant.

The most important case of $M$ is Minkowski spacetime and we 
are interested in the confining solutions $A$ of the SU(3)-Yang-Mills 
equations. The confining solutions were defined in Ref. \cite{Gon01} as the 
spherically symmetric solutions of the Yang-Mills 
equations (1) containing only the components of the 
SU($3$)-field which are Coulomb-like or linear in $r$. Additionally 
we impose the Lorentz condition on the sought solutions. 
The latter condition is necessary for 
quantizing the gauge fields consistently within the framework of perturbation 
theory (see, e. g. Ref. \cite{Ryd85}), so we should impose the given condition 
that can be written
in the form ${\rm div}(A)=0$, where the divergence of the Lie algebra valued
1-form $A=A_\mu dx^\mu=A^a_\mu \lambda_adx^\mu$ is defined by the relation 
(see, e. g., Refs. \cite{{81}})
$${\mathrm{div}}\, {A}=\ast(d\ast{A})=
\frac{1}{\sqrt{\delta}}\partial_\mu(\sqrt{\delta}g^{\mu\nu}
A_\nu)\>.\eqno(\mathrm D.2)$$
It should be emphasized that, from the physical point of view, the Lorentz 
condition reflects the fact of transversality for gluons that arise as quanta 
of SU(3)-Yang-Mills field when quantizing the latter (see, e. g., 
Ref. \cite{Ryd85}).

We shall use the Hodge star operator action on the 
basis differential 2-forms on Minkowski spacetime with local 
coordinates $t, r, \vartheta, \varphi$ in the form
$$\ast(dt\wedge dr)=-r^2\sin\vartheta d\vartheta\wedge d\varphi\>,
\ast(dt\wedge d\vartheta)=\sin\vartheta dr\wedge d\varphi\>,$$
$$\ast(dt\wedge d\varphi)=-\frac{1}{\sin\vartheta}dr\wedge d\vartheta\>,
\ast(dr\wedge d\vartheta)=\sin\vartheta dt\wedge d\varphi\>,$$
$$\ast(dr\wedge d\varphi)=-\frac{1}{\sin\vartheta}dt\wedge d\vartheta\>,
\ast(d\vartheta\wedge d\varphi)=\frac{1}{r^2\sin\vartheta}dt\wedge dr\>,
\eqno(\mathrm D.3)$$
so that on 2-forms $\ast^2=-1$. More details about the Hodge star operator can 
be found in \cite{81}. 

The most general ansatz for a spherically symmetric solution is 
$A=A_t(r)dt+A_r(r)dr+A_\vartheta(r)d\vartheta+A_\varphi(r)d\varphi$. 
But then the Lorentz 
condition (D.2) for the given ansatz gives rise to 
$$\sin{\vartheta}\partial_r(r^2A_r)+
\partial_\vartheta(\sin{\vartheta}A_\vartheta)=0,$$
which yields $A_r=\frac{C}{r^2}-
\frac{\cot{\vartheta}}{r^2}\int A_\vartheta(r)dr$ with a constant matrix $C$. 
But the confining solutions should be spherically symmetric and contain only 
the components which are Coulomb-like or linear in $r$, so one should put 
$C=A_\vartheta(r)=0$. Consequently, 
the ansatz $A=A_t(r)dt+A_\varphi(r)d\varphi$ is the most general 
spherically symmetric one. 

For the latter ansatz we have $F=dA+gA\wedge A=-\partial_rA_tdt\wedge dr+
\partial_rA_\varphi dr\wedge d\varphi+g[A_t,A_\varphi]dt\wedge d\varphi$, 
where $[\cdot,\cdot]$ signifies matrix commutator.
 
Then, according to (D.3), we obtain 
$$\ast F= (r^2\sin{\vartheta})\partial_rA_td\vartheta\wedge d\varphi-
\frac{1}{\sin{\vartheta}}\partial_rA_\varphi dt\wedge d\vartheta-$$ 
$$\frac{g}{\sin{\vartheta}}[A_t,A_\varphi]dr\wedge d\vartheta\>,
\eqno(\mathrm D.4)$$ 
which entails 
$$d\ast F= 
\sin{\vartheta}\partial_r(r^2\partial_rA_t)\,dr\wedge d\vartheta\wedge d\varphi+
\frac{1}{\sin{\vartheta}}\partial_r^2A_\varphi\, dt\wedge dr\wedge d\vartheta
\>,\eqno(\mathrm D.5)$$
while 
$$\ast F\wedge A - A\wedge\ast F=$$
$$\biggl(r^2\sin{\vartheta}[\partial_rA_t,A_t]-
\frac{1}{\sin{\vartheta}}[\partial_rA_\varphi,A_\varphi]\biggr)
\,dt\wedge d\vartheta\wedge d\varphi$$
$$-\frac{g}{\sin{\vartheta}}\biggl([[A_t,A_\varphi],A_t]\,
dt\wedge dr\wedge d\vartheta$$
$$+[[A_t,A_\varphi],A_\varphi]\,
dr\wedge d\vartheta\wedge d\varphi\biggr)
\>.\eqno(\mathrm D.6)$$
Under the circumstances the Yang-Mills equations (D.1) are 
tantamount to the conditions 
$$\partial_r(r^2\partial_rA_t)=-
\frac{g^2}{\sin^2{\vartheta}}[[A_t,A_\varphi],A_\varphi],\eqno(\mathrm D.7)$$
$$\partial_r^2A_\varphi=-
{g^2}[[A_t,A_\varphi],A_t],\eqno(\mathrm D.8)$$
$$r^2\sin{\vartheta}[\partial_rA_t,A_t]-
\frac{1}{\sin{\vartheta}}[\partial_rA_\varphi,A_\varphi]=0.
\eqno(\mathrm D.9)$$
The key equation is (D.7) because the matrices $A_t, A_{\varphi}$ depend on 
merely $r$ and (D.7) can be satisfied only if the matrices 
$A_t=A_t^a\lambda_a$ and $A_\varphi=A_{\varphi}^a\lambda_a$ belong to the 
so-called Cartan subalgebra of the SU(3)-Lie algebra. Let us remind that, by definition, 
a Cartan subalgebra is a maximal abelian subalgebra in 
the corresponding Lie algebra, i. e., the commutator for any two matrices of 
the Cartan subalgebra is equal to zero (see, e.g., Ref. \cite{Bar}). For 
SU(3)-Lie algebra the conforming Cartan subalgebra is generated by the 
Gell-Mann matrices $\lambda_3, \lambda_8$ which are
$$\lambda_3=\pmatrix{1&0&0\cr 0&-1&0\cr 0&0&0\cr}\,,  
  \lambda_8={1\over\sqrt3}\pmatrix{1&0&0\cr 0&1&0\cr 
                   0&0&-2\cr}\,.\eqno(\mathrm D.10)$$
Under the situation we should have $A_t=A_t^3\lambda_3+A_t^8\lambda_8$ and 
$A_\varphi=A_{\varphi}^3\lambda_3+A_{\varphi}^8\lambda_8$, then 
$[A_t,A_\varphi]=0$ and we obtain 
$$\partial_r(r^2\partial_rA_t)=0, \partial_r^2A_\varphi=0, 
\eqno(\mathrm D.11)$$
while (D.9) is identically satisfied and (D.11) gives rise to the 
solution (3) with real constants $a_j, A_j, b_j, B_j$
parametrizing the solution which proves the uniqueness theorem of Section 2 
for the SU(3) Yang-Mills equations. 

The more explicit form of (3) is 
$$A^3_t = [(a_2-a_1)/r+A_1-A_2]/2,\>$$
$$A^8_t =[A_1+A_2-(a_1+a_2)/r]\sqrt{3}/2\>,$$
$$ A^3_\varphi = [(b_1-b_2)r+B_1-B_2]/2,$$
$$ A^8_\varphi= [(b_1+b_2)r+B_1+B_2]\sqrt{3}/2\>.\eqno(\mathrm D.12)$$

Clearly, the obtained results may be extended over all SU($N$)-groups with 
$N\ge2$ and even 
over all semisimple compact Lie groups since for them the corresponding Lie 
algebras possess just the only Cartan subalgebra. Also we can talk about the 
compact non-semisimple groups, for example, U($N$). In the latter case 
additionally to Cartan subalgebra we have centrum consisting from the matrices 
of the form $\alpha I_N$ ($I_N$ is the unit matrix $N\times N$) with arbitrary 
constant $\alpha$. 

The most relevant physical cases are of course U(1)- and SU(3)-ones 
(QED and QCD). In particular, the U(1)-case allows us to build the classical 
model of confinement (see Section 2 and Ref. \cite{GF10}). 

At last, it should also be noted that the 
nontrivial confining solutions obtained exist at any gauge coupling constant 
$g$, i. e. they are essentially {\em nonperturbative} ones.


\end{document}